\documentclass[10pt,letterpaper]{article}
\usepackage[a4paper,
            bindingoffset=0.2in,
            left=1in,
            right=1in,
            top=1in,
            bottom=1in,
            footskip=.25in]{geometry}
\usepackage{tabularx,booktabs}
\usepackage{url}
\usepackage{times}
\usepackage{fullpage}
\usepackage{graphicx}
\usepackage{lipsum}
\newcommand{\junk}[1]{}

\usepackage{multirow}
\usepackage{booktabs}
\usepackage{color}
\usepackage{subfigure}
\usepackage{amsfonts}
\usepackage{amsmath}
\usepackage[dvipsnames]{xcolor}
\usepackage{tikz}
\usepackage{threeparttable}
\usepackage{caption}
\usepackage{lineno}
\usepackage{authblk}

\makeatletter
\newbox\dottedarrow@box
\setbox\dottedarrow@box\hbox
  {%
    \begin{tikzpicture}
      \draw[dotted,->] (0,0) -- (1.5em,0);
    \end{tikzpicture}%
  }
\newcommand*\dottedarrow
  {\relax\ifmmode\expandafter\dottedarrow@m\else\expandafter\dottedarrow@t\fi}
\newcommand*\dottedarrow@t[1][1.5em]
  {\resizebox{#1}{!}{\raisebox{.5ex}{\usebox\dottedarrow@box}}}
\newcommand*\dottedarrow@m[1][]
  {%
    \if\relax\detokenize{#1}\relax
      \mathchoice% values are trial and error based\ldots
        {\dottedarrow@t}
        {\dottedarrow@t}
        {\dottedarrow@t[1.1em]}
        {\dottedarrow@t[0.9em]}%
    \else
      \dottedarrow@t[#1]%
    \fi
  }
\makeatother

\usepackage[numbers,sort&compress]{natbib}

\usepackage{hyperref}
\hypersetup{hypertex=true,
            colorlinks=true,
            linkcolor=blue,
            anchorcolor=blue,
            citecolor=blue}

\newcommand{\forceindent}{\leavevmode{\parindent=1.5em\indent}}

\usepackage{xr}
\begin{document}
\thispagestyle{empty}

\title{Interpretable RNA Foundation Model from Unannotated Data for Highly Accurate RNA Structure and Function Predictions}
% \author
% {
% \centering
% authors
% % Master$^{1}$ \and Yoda$^{2}$ \and Doctor$^{1}$\footnote{Corresponding Author. Email: thirteenth@doctorwho.com} 
% % \\
% % 		$^{1}$Dept of Dalek Affairs, Time Lord Academy, 42424 Gallifrey\\
% % 	   	$^{2}$Dept of Light Saber Engineering, Jedi Academy, 12345, Coruscant\\	
% }

% \author{authors}
\author[1]{Jiayang Chen$^\dagger$}
\author[1]{Zhihang Hu$^\dagger$}
%\author[2,3]{Siqi Sun$^{\dagger*}$}
\author[2,3]{Siqi Sun\thanks{Corresponding Author. Email: liyu@cse.cuhk.edu.hk and siqisun@fudan.edu.cn.}\thanks{Equal first authorship.}}
\author[1]{Qingxiong Tan}
\author[1,4]{Yixuan Wang}
\author[1,5]{Qinze Yu}
\author[1]{Licheng Zong}
\author[1]{Liang Hong}
\author[1]{Jin Xiao}
\author[1,9]{Tao Shen}
\author[1]{Irwin King}
% \author[6,7,8]{James J. Collins}
\author[1,6,7,8,10]{Yu Li$^*$}
\affil[1]{\small Department of Computer Science and Engineering, The Chinese University of Hong Kong, Hong Kong SAR, China}
\affil[2]{\small Research Institute of Intelligent Complex Systems, Fudan University, Shanghai, China}
\affil[3]{\small Shanghai AI Laboratory, Shanghai, China}
\affil[4]{\small Harbin Institute of Technology, China}
\affil[5]{\small University of Electronic Science and Technology of China, China}

\affil[6]{\small Institute for Medical Engineering and Science, Massachusetts Institute of Technology, Cambridge, MA, USA}
\affil[7]{\small Wyss Institute for Biologically Inspired Engineering, Harvard University, Boston, MA, USA}
\affil[8]{\small Broad Institute of MIT and Harvard, Cambridge, MA, USA}
\affil[9]{\small Zelixir Biotech, Shanghai, China}
\affil[10]{\small The CUHK Shenzhen Research Institute, Hi-Tech Park, Nanshan, Shenzhen, China}

\date{}

\maketitle
\begin{abstract}
  % Inspired by the success of the application of NLP in the protein field, we train an RNA language model, RNA-FM, on 26 million sequences from the RNAcentral dataset by unsupervised learning. 
Non-coding RNA structure and function are essential to understanding various biological processes, such as cell signaling, gene expression, and post-transcriptional regulations. These are all among the core problems in the RNA field.
%Similar to protein structure prediction,  
%RNA structure and function prediction , and its performance could be further improved.
With the rapid growth of sequencing technology, we have accumulated a massive amount of unannotated RNA sequences. On the other hand, expensive experimental observatory results in only limited numbers of annotated data and 3D structures. Hence, it is still challenging to design computational methods for predicting their structures and functions. The lack of annotated data and systematic study causes inferior performance.
% utilizing them to develop practical models.
To resolve the issue, we propose a novel RNA foundation model (RNA-FM) to take advantage of all the 23 million non-coding RNA sequences through self-supervised learning. Within this approach, we discover that the pre-trained RNA-FM could infer sequential and evolutionary information of non-coding RNAs without using any labels. Furthermore, we demonstrate RNA-FM's effectiveness by applying it to the downstream secondary/3D structure prediction, SARS-CoV-2 genome structure and evolution prediction, protein-RNA binding preference modeling, and gene expression regulation modeling. The comprehensive experiments show that the proposed method improves the RNA structural and functional modelling results significantly and consistently. Despite only being trained with unlabelled data, RNA-FM can serve as the foundational model for the field.

\end{abstract}

\section*{Introduction}
  
% \lipsum[1-3]

%% the main contents:

% \textbf{describe the basic conception importance of of RNA:}
% \textbf{describe the importance in the RNA structure and function prediction:}
% According to the basic dogma of the molecular biology, millions of gene sites are transcribed to produce Ribonucleic acid (RNA) transcripts (namely DNA → RNA), which are then translated to produce different types of protein sequences (namely RNA → protein) so as to maintain the normal functions of human bodies  \cite{chen2019computational,li2015statistics,dreyfuss1996transcript,mcknight1982transcriptional}.
% Such central dogma indicates that 
% \textcolor{blue}{Ash}
\forceindent RNA plays an important role in performing various types of biological functions, such as cell signaling, gene expression, and post-transcriptional regulations \cite{miao2017rna,caprara2000rna,atkins2011rna}. Determination of RNA structure or type is also an essential part of RNA-based therapeutics, including mRNA vaccines, RNA interference and CRISPR-based therapeutics \cite{li2020strategies, bora2012rna, pardi2018mrna}. Among all RNA transcripts, about 5\% served as mRNAs responsible for protein coding, while the substantial remaining portion is non-coding RNAs (ncRNAs) \cite{ chen2019computational, wang2011molecular}. Particularly, these ncRNAs must sustain specific structures to conduct corresponding biological functions. Different ncRNAs, such as small nuclear ribonucleoproteins (snRNPs), ribosomes, microRNAs, small nucleolar ribonucleoproteins (snoRNPs), long ncRNAs, and telomerase, also interact with
proteins to form stable RNA-protein complexes to perform specific functions \citep{cech2014noncoding, miao2017rna}.
Accurately modelling ncRNAs structures could help understand their performed functions and various biological processes. Despite a large number of ncRNA sequences, few of their structures and functions are known \citep{townshend2021geometric,yao2019cellular}. Traditionally, RNA three-dimensional (3D) structures assessed by experimental approaches, including nuclear magnetic resonance, X-ray crystallography and cryogenic electron microscopy, are expensive and time-consuming. Therefore, computational approaches are developed and applied to bridge the gap.

Regarding the RNA structure prediction, most existing approaches focus on the RNA secondary structure prediction, which could further be divided into three categories: thermodynamic methods, alignment-based methods, and deep learning (DL)-based methods. Foundational works can be traced back to the 1980s, when thermodynamic parameters-based methods were proposed \citep{zuker1984rna, waterman1978rna, rivas2013four} to predict RNA secondary structures. Nowadays, combined with dynamic programming, these approaches such as Vienna RNA \citep{ stadler2011viennarna,hofacker1994fast}, Mfold/UNAFold \citep{markham2008unafold,zuker2003mfold}, LinearFold \citep{huang2019linearfold}, and RNAstructure \citep{mathews1998updated,reuter2010rnastructure} are still widely used since they can incorporate a large volume of folding features (together with standard base pairs) to estimate the model parameters.
Thermodynamic-based methods are later improved by considering parameters of local structures, including experimental parameters (such as RNAstructure \cite{reuter2010rnastructure}, RNAfold \cite{lorenz2011viennarna}, RNAshapes \cite{janssen2015rna}) and machine learning parameters (such as ContextFold \cite{zakov2011rich}, CONTRAfold \cite{do2006contrafold}, CentroidFold \cite{sato2009centroidfold}). 
Though many currently used approaches fall into this category, their overall performance seems to hit the plateau. These methods usually do not consider all base pairs obtained from tertiary interactions \cite{nowakowski1997rna,rivas2013four,sato2009centroidfold}, which may miss essential information in their predictions.

To overcome the above problem, alignment-based methods built upon comparative sequence analysis are thus designed to determine the vital base pairs among homologous sequences \cite{singh2019rna}.
With the help of sufficient homologous sequences and their alignments, these alignment-based methods achieve excellent performance in predictions \cite{gutell2002accuracy,singh2019rna}. 
However, new issues arose from the limited number of known RNA families, given Rfam only contains several thousand RNA families. Since RNA is much less conserved, such property restricts the further improvement of alignment-based methods theoretically compared to the vast number of protein families.
On the other hand, with more RNA data available, several deep learning approaches have been recently developed in the community to improve the accuracy of RNA secondary structure prediction. For example, SPOT-RNA \cite{singh2019rna}, E2Efold \cite{chen2020rna}, MXfold2 \cite{sato2021rna}, and UFold \cite{fu2021ufold} are shown to be able to improve the prediction accuracy significantly on different datasets.
Nevertheless, the generalization capability of such DL-based methods still remains a problem, as the model architecture is explicitly designed for corresponding tasks and cannot generalize well to unknown RNA types \cite{chen2020rna}.

In contrast to secondary structure prediction, the modelling of RNA 3D structure is still under-explored due to the lack of 3D structure data.
In fact, computational optimization combined with deep learning methods may serve as an alternative way to solve the 3D problem. While there exist methods to optimize 3D structure with the minimum energy \citep{xiong2021pairing} given 2D information, deep learning methods could be utilized to solve the downgraded 2D problem (secondary structure, distance structure). 
For example, RNAcontact \cite{sun2021rna} could predict 3D closeness between base pairs by the deep residual neural networks. ARES, which consists of many processing layers, was proposed to score the predicted RNA structures \citep{townshend2021geometric}.
Despite the above attempts so far, there are no end-to-end DL-based methods that could generate RNA 3D structure directly. The lack of annotated 3D data is one major obstacle. 
% Other barriers, including designation or complexity, could be overcome gradually.

In addition to RNA structure, understanding its function is also vital. In particular, predicting the interaction between RNAs and proteins \cite{lam2019deep,sun2021predicting} could help to understand gene expression regulation. We could utilize either data from existing databases or data generated by biological experiments that classified RNAs into several functional groups for this application. With these extensive hand-crafted labels, DL-based approaches are then proposed to learn the underlying distribution of RNAs in different functional groups. Therefore, the corresponding functional group could be predicted given query RNA sequences. For instance, a deep discriminative neural network was developed to distinguish RNAs that can bind to specific RNA-binding proteins from the non-binding ones \cite{sun2021predicting}.
Furthermore, 1D CNN is designed to predict the protein expression levels regulated by human 5' UTRs from RNA sequence \cite{sample2019human}.

As the previous computational approaches rely heavily on the label information of RNA sequences, they all share the limitation of generalization. A model needs to be deliberately designed and tuned on specific tasks, yet challenging to transfer to other related studies. For instance, even E2Efold, a promising RNA secondary structure prediction model, performs well on dominant RNA types, including tRNAs and 5S rRNAs. Its performance on unknown RNA types degrades significantly. Also, models built upon RNA-protein interactions cannot gain accurate predictions on the UTR regulation. They are unfriendly to biological researchers because much effort will be needed to retrain specific deep learning based models. 
% This challenge causes internal friction beyond the community.

\begin{figure}[!t]
\centering
\includegraphics[width=0.98\textwidth]{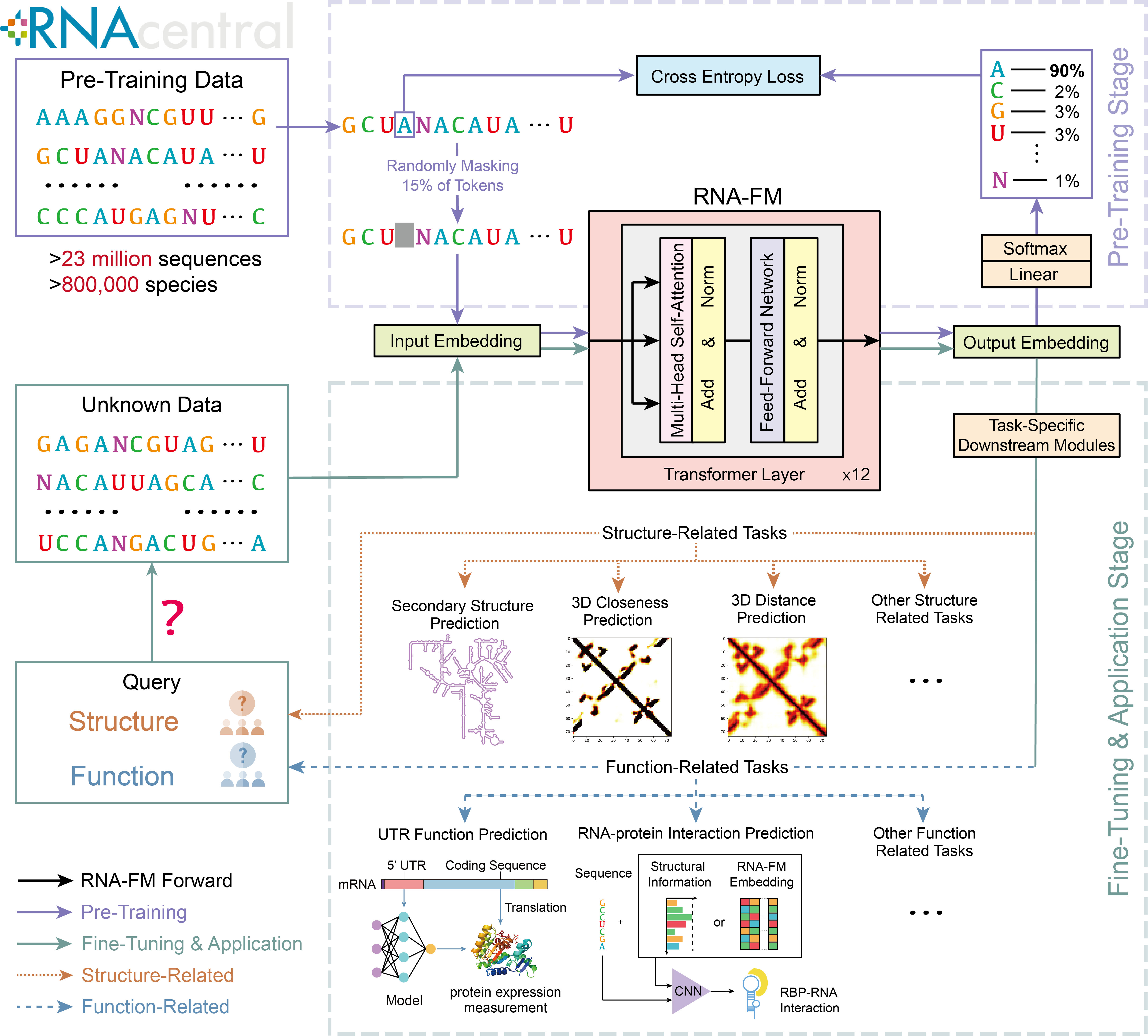} 
\caption{
\textbf{Overview of RNA foundation model (RNA-FM) design and applications}. RNA-FM consists of 12 transformer layers. In the pre-training stage, the RNA-FM is trained with 23 million sequences from the RNAcentral database via self-supervised learning, \textit{i.e.}, reconstructing the masked tokens from the sequence alone. 
Thus, we could obtain effective feature representations of the query RNA sequences without using any labels.
In the fine-tuning stage, the representations from RNA-FM could significantly and consistently improve the performance on both structure-related and function-related applications with simple task-specific prediction modules.
} 
\label{Fig.overview}
\end{figure}

Notice that virtually every method omits the millions of unannotated RNA sequences and only utilizes the small annotated dataset with at most 30K sequences. 
To overcome the above generalization limitation, we suggest taking advantage of the enormous unannotated RNA sequence data, which contains the evolutionary information of RNA sequences. We propose a novel RNA foundation model (RNA-FM), as shown in Figure \ref{Fig.overview}, implying `ONE-FOR-ALL', with which various RNA-related downstream applications can be conducted via replacing its predicting layers only. The model is trained in a self-supervised manner that differs from all the previous RNA DL-based approaches. To the best of our knowledge, we are the first that attempt to extract meaningful RNA representations from unlabeled data and improve multiple downstream applications simultaneously. Precisely, the model is presented in two phases: pre-training and fine-tuning. Throughout the task-agnostic pre-training stage, the proposed RNA-FM is trained with 23 million ncRNA sequences from the RNAcentral database using self-supervised learning. Eventually, RNA-FM learns the sequential distribution and pattern that could potentially capture the underlying structural and functional information. Then, during the task-specific fine-tuning stage, the pre-trained RNA-FM model either generates sequence embeddings (features) that fit into downstream modules or is fine-tuned with a lightweight prediction layer.
With the powerful representation learned from unannotated ncRNA data, RNA-FM significantly improves the performance across a broad range of RNA structure/function-related prediction applications with minor modifications to the model architectures. In terms of RNA secondary structure prediction, RNA-FM outperforms LinearFold by up to 30\% and SPOT-RNA by up to 20\% in terms of F1 score on cross-dataset validation. Even on the low-redundant dataset, which is rather challenging for previous methods, RNA-FM still outperforms SPOT-RNA by up to 7.5\% and UFold by 4\% in terms of F1 score. As for 3D closeness prediction, a single model built upon RNA-FM can even exceed an ensemble method with 100 models by 30\% regarding the long-range top precision.
In addition to working well on the benchmark datasets, RNA-FM can model the regulatory elements in the SARS-CoV-2 genome and potentially illustrates the evolution of the virus variants.
Furthermore, the embedding from RNA-FM can achieve comparable performance with the \textit{in vivo} secondary structure feature on protein-RNA interaction prediction, even though RNA-FM is trained on unannotated RNA sequences only. 
All of these performance improvements suggest that the proposed large-scale pre-trained foundation model has implicitly captured both structural and functional information from RNA sequences alone.

We summarize the main contributions of this work as follows:
\begin{itemize}
\item We propose the first RNA foundation model, RNA-FM, delivering rich representations for the ncRNA universe. In addition, we provide a web server such that the community can access the pre-trained model and its powerful representations. The server can be accessed with this link: \href{https://proj.cse.cuhk.edu.hk/rnafm/}{https://proj.cse.cuhk.edu.hk/rnafm/} The code and weights are available at \href{https://github.com/ml4bio/RNA-FM}{https://github.com/ml4bio/RNA-FM}. 

\item RNA-FM can produce interpretable RNA representations, which contain evolutionary information. Such embeddings can be used to infer the evolutionary trend of lncRNAs and SARS-CoV-2 variants.

\item The representations generated from RNA-FM can substantially improve the performance on multiple ncRNA structure/function prediction applications, with a desirable generalization property to the regulatory UTR regions of mRNAs and SARS-CoV-2.

\end{itemize}

\section*{Results}
\label{result}

\begin{figure}[!t]
\centering
\includegraphics[width=0.90\textwidth]{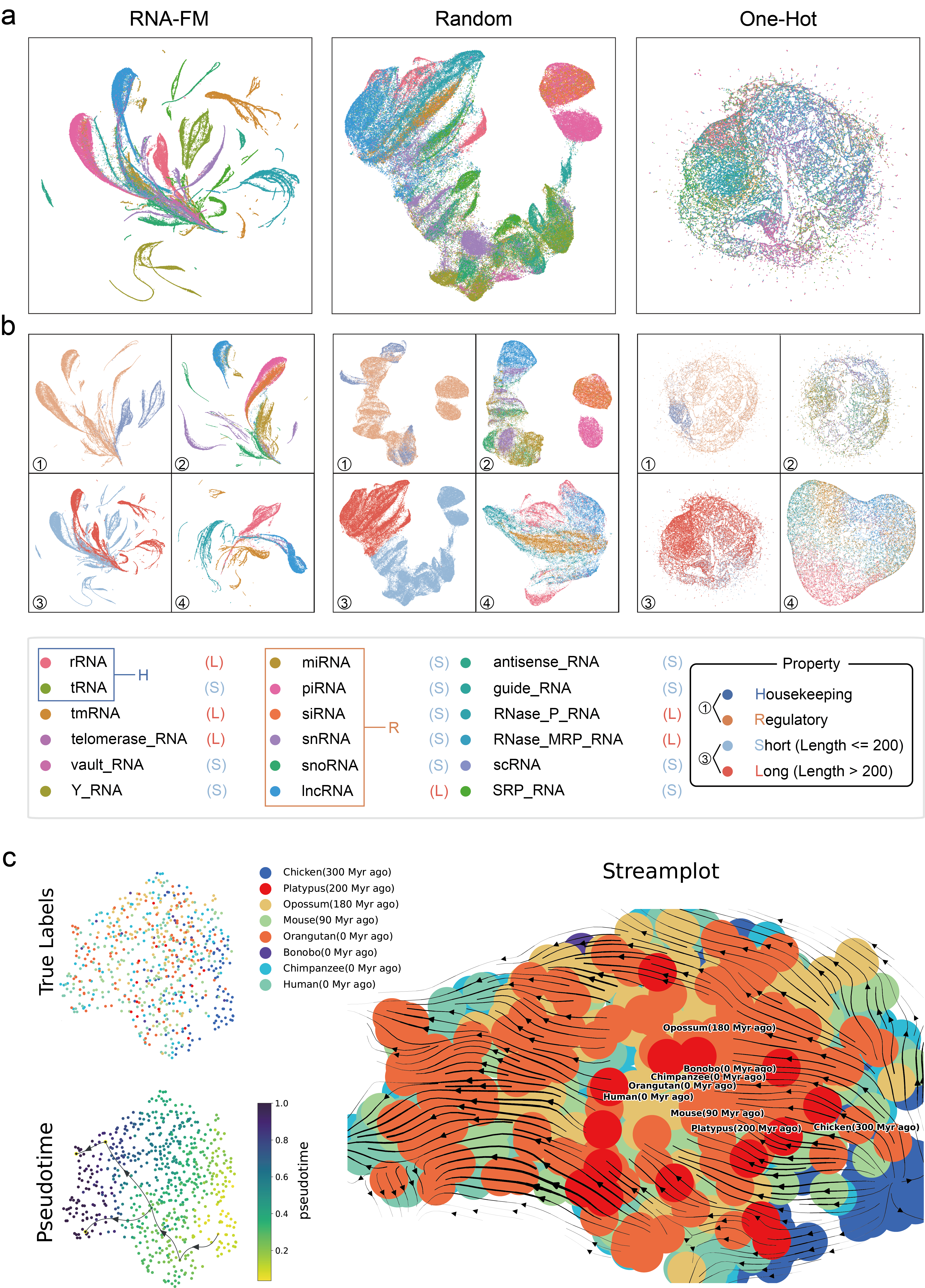} 
\caption{\textbf{RNA-FM encodes multi-scale patterns.} \textbf{a.} RNA Atlas with all ncRNA types in RNAcentral100 using different embedding ways. 
\textbf{b.(1)} Embedding projections of housekeeping RNA (rRNA, tRNA) and regulatory RNA (lncRNA, snoRNA, miRNA, siRNA, snRNA, piRNA).
\textbf{b.(2)} Detailed distribution of the regulatory RNAs. 
\textbf{b.(3)} Embedding projections of long ncRNA (sequence length $>$200, rRNA, tmRNA, \textit{etc.}) and small ncRNA (sequence length $<$200, tRNA, sncRNA, \textit{etc.}). 
\textbf{b.(4)} Detailed distribution of long RNAs (most of the small RNAs have been analyzed in b.(2)). 
\textbf{c.} Trajectory inference of lncRNA evolutionary trend based on the RNA-FM embeddings.
}
\label{Fig.visual}
\end{figure}

\paragraph{Learning from large-scale unlabeled non-coding RNA sequences.}
As shown in Figure \ref{Fig.overview}, in order to take advantage of a massive amount of unlabeled ncRNA data and avoid relying on label information, we propose our RNA foundation model (RNA-FM) based on the BERT \cite{devlin2018bert} language model architecture. It is built upon 12 transformer-based bidirectional encoder blocks and trained on 23 million sequences from the RNAcentral database in a self-supervised manner.
After training, RNA-FM produces a $L\times640$ embedding matrix for each RNA sequence with length $L$. These embeddings are expected to contain rich information within the ncRNA universe. We verified the effectiveness of RNA-FM on various applications. Firstly, to investigate what has been learned by the model and the physical meaning of the model outputs, we analyze the derived embeddings directly and examine how ncRNAs of similar function and structure gather in a 2-dimensional plane, resulting in the RNA Atlas in Figure \ref{Fig.visual}. Also, the embedding from RNA-FM can be used to infer the long non-coding RNA (lncRNA) evolutionary trend, which suggests that the evolutionary information has been learned by our model implicitly. Furthermore, models using RNA-FM embeddings could improve over state-of-the-art approaches consistently on various structural-related and functional-related downstream prediction problems, including both SARS-CoV-2 study and gene expression regulation modeling.

\paragraph{RNA-FM learns multi-dimension biological information of RNA universe.}

To demystify what has been learned by the large-scale RNA-FM model from the million-scale data and the physical meaning of the model outputs, we take a closer look into RNA biological information contained in the RNA-FM embedding, including its structural/functional properties and evolutionary information. Such analysis shows the interpretability of the RNA-FM model.

\textbf{RNA functions and structures} vary across different RNA types, and RNA-FM is expected to encode these rich properties within our generated embeddings from pure RNA sequences. We build RNA Atlas by employing the proposed encoder on the known RNA universe to analyze their patterns. Next, UMAP \cite{mcinnes2018umap} is applied to reduce the dimension of embeddings and project them into a 2-dimensional plane. To compare and visualize how the learning process shapes the representations, we take both generated representations before and after model pre-training into consideration. Furthermore, random initialized RNA-FM (Random) and one-hot encoding (One-Hot) are also introduced for visualization purposes.
% \jiayang{intraclass}
The results are shown in Figure \ref{Fig.visual}a. In the pre-trained RNA-FM embedding space (left), visual inspection reveals that RNA types are organized by structure and function properties with clear boundaries between clusters. In contrast, the projection of the Random model (middle) presents some extremely vague clustering structure; while projections of One-Hot encoding (right) are barely distinguishable with no apparent structure information, implying that RNA-FM has learned structural or functional information beyond their primary structure, such that instances with similar properties are grouped.

% \jiayang{different view} 
We take it a step further to discuss ncRNAs from different categorical views, such as housekeeping ncRNAs and regulatory ncRNAs, long ncRNAs ($>$200 nucleotides), and small ncRNAs ($<=$200 nucleotides). In Figures \ref{Fig.visual}b(1) and b(3), RNA-FM well discriminates housekeeping and regulatory categories but struggles to deal with short and long ncRNAs. This might suggest that RNA-FM encoding emphasizes more on structural or functional similarity rather than length since RNAs with different lengths could share the same functions and RNAs with similar lengths could differ significantly.
% \jiayang{interclass \& length effect}
In Figures \ref{Fig.visual}b(2) and b(4), when we look closer into some sub-classes within a limited length range (less or greater than 200) in the RNA-FM part. These RNA embeddings aggregate or separate according to the similarity of their structures and functions; it proves again that RNA-FM establishes the RNA Atlas by recognizing the structures and functions of RNAs rather than their length. %discover that interclass patterns of embeddings are the same as the above intraclass ones for both small and long RNAs. 

\textbf{RNA evolutionary information} is also explored in our studies. We apply trajectory inference (pseudotemporal ordering) \cite{saelens2019comparison}, which is commonly used in single-cell transcriptomics, to a subset of lncRNA with their RNA-FM embedding as input. RNAs in the subset can be classified according to different types of species. Here we obtain their evolutionary relationship from an evolutionary study of lncRNA repertoires and expression patterns \cite{necsulea2014evolution}. We first generate RNA-FM embeddings for the lncRNA subset, then trajectory inference is carried out via VIA \cite{stassen2021generalized} and the stream-plot is shown in Figure \ref{Fig.visual}c. We discover that although it is hard for RNA-FM to distinguish these RNAs into different species, the embeddings are able to present a roughly accurate evolutionary trend of different species corresponding to their ground-truth timeline. The result is surprising because we do not include such evolutionary features during training and only use the pure RNA sequences. This result testified that RNA-FM deeply mined the implicit genetic message and encoded the outputs with evolutionary information.

% We first directly analyse the embedding space formed by our RNA-FM and generate RNA Atlas based on the RNAcentral. To further evaluate the effectiveness of our RNA-FM, we apply it to a variety of downstream tasks, including structure-related and function-related tasks. As a result, the RNA-FM can improve almost 20\% over SOTAs in the structure prediction tasks. In addition, the RNA-FM can also contribute to function-related tasks to some extent.

% Unsupervised language models are usually deemed to encode the obvious underlying patterns from vast amounts of sequences. To find whether our RNA-FM can capture RNA sequence patterns and understand how the pre-training shapes the representations, we compare the changes of the embedding spaces formed by RNA-FM on RNAcentral100 before and after the pre-training phase. The analyses are conducted from two scales, including sequence-level and gram-level.

% \paragraph{Embeddings encode RNA Sequence Type}

% In order to compare the shaped representations before and after the pre-training phase, we project the embedding of them into two-dimensions using UMAP.

% \paragraph{Embeddings encode RNA N-gram pattern} supp

\paragraph{RNA-FM benefits both RNA secondary structure prediction and 3D modelling results.} Structure understanding is always the key among various RNA-related applications since RNA structure usually determines its function. However, only a tiny fraction (\textless0.001\%) of the structure-known ncRNAs has been determined by experiments \cite{zhao2021review} due to the high cost of wet-lab experiments and RNA structural instability. To tackle this problem, more and more computational approaches \cite{singh2019rna, chen2020rna,sato2021rna,fu2021ufold} have been proposed for RNA structure prediction. We investigate RNA-FM's performance on several structure prediction tasks, including secondary structure prediction, 3D closeness prediction, and RNA map distance prediction. We also try to perform 3D reconstruction and prediction beyond the predicted secondary structures.

\textbf{RNA secondary structure} can be rapidly formed from its primary sequence by pairing bases with hydrogen bonds. Secondary structure is much more stable and more accessible in cells than its tertiary form, making it an essential role for the high-order structure prediction or even function prediction \cite{zhao2021review}. This section performs a comprehensive comparison of RNA-FM and other popular RNA secondary structure prediction methods, as well as a head-to-head comparison with one of the SOTA methods, UFold \cite{fu2021ufold}.

%\textit{Evaluation}
We conduct experiments on several benchmarks commonly used in E2Efold, SPOT-RNA and UFold.
(1) RNAStralign \cite{tan2017turbofold}, which consists of 37149 structures from 8 RNA types, is one of the most comprehensive collections of RNA structures in the field.
(2) ArchiveII \cite{sloma2016exact}, which consists of 3975 RNA structures from 10 RNA types, is also a widely-used benchmark dataset for many classical RNA folding methods.
(3) bpRNA-1m \cite{singh2019rna}. The dataset is preprocessed by removing sequence similarity with 80\% sequence-identity cut-off and restricting their maximum sequence length below 500. The preprocessed dataset contains 13,419 sequences and is randomly split into 10,814 RNAs for training (TR0), 1300 for validation (VL0), and 1,305 for testing (TS0).
% (4) PDB \cite{singh2019rna}. The dataset for transfer learning is obtained by downloading high-resolution ($<3.5$\r{A}) RNAs from PDB on March 2, 2019. After applying CD-HIT-EST with 80\% cut-off and BLAST-N, we finally obtain 120, 30, and 67 RNAs for training (TR1), validation (VL1), and independent test (TS1), respectively.
For evaluation and testing, we take the usage of Ufold \cite{fu2021ufold} data and make a fair comparison. The well-trained model is evaluated on ArchiveII600 (a subset with a length less than 600) and TS0.

%Here, we only their offered subset of length less than 600 due to the input limitation of our model and regard them as RNAStralign600 and ArchiveII600. The RNA-FM and ResNet32 are trained on the RNAStralign600 and directly evaluated on ArchiveII600 without re-training.

\begin{table}[t]
\centering
\caption{\textbf{RNA secondary structure prediction performance.} Our method beats the other 12 SOTA methods on the two datasets across all evaluation criteria except for being slightly behind Ufold on the recall score. RNA-FM is not specific to RNA secondary structure prediction. However, it has learned rich information about the secondary structure and done a much better job in RNA secondary structure prediction than other models.}
\label{Tab.ss}
\begin{threeparttable}
\begin{tabular}{ccccccc} 
\toprule
\multirow{2}{*}{Method}     & \multicolumn{3}{c}{ArchiveII600 (3911)} & \multicolumn{3}{c}{bpRNA TS0 (1305)}  \\ 
\cmidrule(lr){2-4}\cmidrule(lr){5-7}
                            & Pre\tnote{a}   & Rec   & F1s                     & Pre   & Rec   & F1s                   \\ 
%\midrule
% RNA-FM (All dataset)        & 0.918 & 0.942 & 0.928                   & 0.700 & 0.712 & 0.696                 \\ 
\midrule
RNA-FM & \textbf{0.936} & \textbf{0.951} & \textbf{0.941}                   & \textbf{0.718} & 0.713 & \textbf{0.704}                \\
%RNA-FM (Pretrain + Finetune) & \textbf{0.924} & \textbf{0.958} & \textbf{0.939}                   & \textbf{0.684} & 0.725 & \textbf{0.694}                 \\
% RNA-FM (Pretrain + Feature)  & 0.873 & 0.913 & 0.890                   & 0.708 & 0.609 & 0.639                 \\
% RNA-FM (Random + Finetune)   & 0.861 & 0.910 & 0.882                   & 0.668 & 0.578 & 0.600                 \\
% RNA-FM (Random + Feature)    & 0.860 & 0.910 & 0.882                   & 0.632 & 0.572 & 0.577                 \\ 
\midrule
UFold                       & 0.890 & 0.926 & 0.905                   & 0.607 & \textbf{0.741} & 0.654                 \\
E2Efold                     & 0.738 & 0.665 & 0.690                   & 0.140 & 0.129 & 0.130                 \\
LinearFold                  & 0.641 & 0.617 & 0.621                   & 0.561 & 0.581 & 0.550                 \\
Mfold                       & 0.428 & 0.383 & 0.401                   & 0.501 & 0.627 & 0.538                 \\
RNAstructure                & 0.563 & 0.615 & 0.585                   & 0.494 & 0.622 & 0.533                 \\
RNAfold                     & 0.565 & 0.627 & 0.592                   & 0.494 & 0.631 & 0.536                 \\
CONTRAfold                  & 0.607 & 0.679 & 0.638                   & 0.528 & 0.655 & 0.567                 \\
SPOT-RNA                    & 0.743 & 0.726 & 0.711 
& 0.594 & 0.693 & 0.619                 \\
RNAsoft	                    & 0.665 & 0.594 & 0.622
 & 0.497 & 0.626 & 0.535                 \\
MXfold2	                    & 0.788 & 0.760 & 0.768
 & 0.519 & 0.646 & 0.558                 \\
Contextfold	                & 0.873 & 0.821 & 0.842 	
 & 0.529 & 0.607 & 0.546                 \\
Eternafold	                & 0.667 & 0.622 & 0.636 	
 & 0.516 & 0.666 & 0.563                 \\

\bottomrule
\end{tabular}
\begin{tablenotes}
        \footnotesize
        \item[a] Pre, Rec, and F1s are the macro averages of the precision, recall and F1-score, respectively.  %此处加入注释*信息
      \end{tablenotes}
\end{threeparttable}
\end{table}

Table \ref{Tab.ss} presents the accuracy of proposed RNA-FM and other advanced approaches \cite{fu2021ufold, huang2019linearfold,zuker2003mfold, reuter2010rnastructure,chen2020rna,singh2019rna,do2006contrafold,sato2021rna,lorenz2011viennarna,zakov2011rich,andronescu2003rnasoft, wayment2020rna} to secondary structure prediction.
RNA-FM outperforms all the other approaches concerning almost all metrics. In addition, RNA-FM far exceeds SPOT-RNA by a total of \textbf{22.8} points and \textbf{7.5} points on the ArchieveII600 and TS0, and distinctly higher than the SOTA (UFold) by a total of \textbf{3.4} points and \textbf{4.0} points on the ArchieveII600 and TS0, respectively, despite UFold also utilizes prior knowledge to model the probability of pairing. The superior performance demonstrates the advantages of underlying structural information encoded by RNA-FM.

Furthermore, we also conduct a head-to-head comparison of RNA-FM with UFold on ArchiveII600. Appendix Figure \ref{Fig.UFold1}(a) shows the F1 score distribution across all samples in ArchieveII600, comparing the RNA-FM with UFold, corresponding to the y-axis and x-axis in the scatter plot, respectively. The RNA-FM matches or exceeds the UFold on \textbf{85.5\%} of instances of all RNA types in the form of most points over the diagonal. We also explore the F1 score on different lengths of the input sequence, as shown in Appendix Figure \ref{Fig.UFold1}(b). Regardless of the length of input RNA sequences, RNA-FM always outperforms UFold, especially when the RNA length is over 150, suggesting that our model better predicts the secondary structure of longer RNA sequences.
Appendix Figure \ref{Fig.UFold2} presents the binary maps predicted by the model with a threshold of 0.5 and a graph view of the secondary structure predictions of two randomly selected examples. The probability maps from RNA-FM (second column) are more robust, less noisy, and much closer to the ground truth (first column) compared to those of UFold (third column). Regarding the graph-view converted from the binarized probability map, RNA-FM also generates secondary structures more similar to the ground truth than UFold.

\textbf{RNA 3D closeness} indicates that arbitrary two bases have tertiary interaction if their distance is under a certain threshold, which originates from the “contact” concept in the protein field.
Although secondary structure can reveal parts of the relationship between base pairs of RNA, it is merely a prior result and usually a constraint applied to the subsequent structure modelling. To obtain more precise structures, researchers propose many informative and challenging tasks for generating more strict constraints for downstream modelling methods.
RNA 3D closeness utilizes a 2D matrix to represent pairwise tertiary inter-nucleotide interactions rather than the 2D flat relationship in secondary structure. The distance is defined as the minimal atomic distance of arbitrary bases, and the threshold is set as 8\r{A}.

%\textit{Evaluation} 
We select the benchmark datasets used by RNAcontact \cite{sun2021rna}, which is constructed based on a set of non-redundant RNA 3D structures from Leontis and Zirbel (2012) (Version 3.99, 2019-11-06), containing 1786 entries with resolution $<4$\r{A} initially. Following preprocessing steps \cite{sun2021rna}, we remove sequences with length $<32nt$ or $>1000nt$, with redundancy over $80\%$ as well as with too few positive points ($<5$). Finally, 221 sequences left are used for training (which we denoted as TR221), and 80 sequences for testing (denoted as TE80). The ground truth is computed from their PDB files following the steps above. The other features involved in the RNAcontact pipeline include the covariance of MSA and the secondary structure predicted by the PETfold \cite{seemann2008unifying} based on MSA.
Appendix Figure \ref{Fig.rnacontact} compares RNA 3D closeness prediction performance of RNAcontact and RNA-FMs with different initialization strategies and training schemes. Table \ref{Tab.rnacontact} then presents the long-range top precisions of different models on the TE80 in detail. We focus on switching input representation features on the same ResNet32 architecture to achieve fair comparisons. Note that we only train ResNet32 once instead of averaging an ensemble of 100 models mentioned in RNAcontact \cite{sun2021rna}. A simple ResNet32 with RNA-FM embeddings achieves SOTA in all aspects and a great improvement over RNAcontact. Then we find that the long-range Top-L precision \cite{wang2017accurate} improves \textbf{7} points when using our RNA-RM embeddings rather than using the covariance and PETfold prediction results. To pursue better performance on such a small dataset, we also adapt the transfer learning by initializing ResNet32 with the parameters pre-trained on bpRNA-1m in the above secondary prediction task. The transfer learning improves the performance significantly by another \textbf{20} points. Obviously, for the small-scale dataset, the pre-trained parameter is critical for both the backbone and downstream model. In addition, The long-range Top-L precision of the model with RNA-FM embedding is always higher than those with MSA covariances or the PETfold secondary structure, which indicates our embeddings from pure sequences present much more useful information than these features from MSA and further eliminate the time-consuming multiple sequence alignment generation step.

\begin{table}
\centering
\caption{\textbf{RNA 3D closeness prediction on the RNAcontact Test80 dataset (long-range top-precision).} The first row shows the results of RNAcontact with the sequence encoding as input (ensemble result of 100 models). The rest rows contain the results predicted by ResNet32 models with different feature inputs. The model with the RNA-FM embedding has already outperformed models with all the other features significantly (20\% performance improvement on Top-L precision over RNAcontact), and its performance can be further boosted dramatically by transfer learning (33\% performance improvement Top-L precision over RNAcontact). Unlike other features generated from MSA data, RNA-FM embedding is obtained from pure sequences, eliminating the time-consuming step of doing multiple sequence alignment.}
\label{Tab.rnacontact}
\begin{threeparttable}
\begin{tabular}{ccccccccc} 
\toprule
\multirow{2}{*}{Features} & \multirow{2}{*}{Source} & \multirow{2}{*}{Model} &
\multicolumn{4}{c}{Long-Range Top Precision}  \\ 
\cmidrule(lr){4-7}
&       &       & L/10 & L/5  & L/2  & L/1    \\ 
\midrule
%\multirow{3}{*}{RNAcontact} 
Seq\tnote{a}     & Seq   & RNAcontact (100 ensemble)   & 0.48 & 0.45 & 0.40 & 0.33   \\
%& Cov       & MSA   & 100   & No   & 0.81 & 0.80 & 0.73 & 0.59   \\
%& Cov + SS  & MSA   & 100   & No   & 0.89 & 0.88 & 0.81 & 0.66   \\ 
\midrule
%\multirow{4}{*}{ResNet32}   
Cov      & MSA   & \multirow{3}{*}{ResNet32 (random)}  & 0.57 & 0.54 & 0.45 & 0.34   \\
Cov + SS\tnote{b}   & MSA   &  & 0.62 & 0.61 & 0.54 & 0.46   \\
RNA-FM\tnote{c} & Seq   &    & \underline{0.68} & \underline{0.66} & \underline{0.62} & \underline{0.53}   \\
\midrule
RNA-FM & Seq    & ResNet32 (transfer)    & \textbf{0.88} & \textbf{0.85} & \textbf{0.79} & \textbf{0.66}   \\
\bottomrule

\end{tabular}
\begin{tablenotes}
        \footnotesize
        \item[a] \textit{Seq} means sequence one-hot encoding
        \item[b] \textit{Cov} means MSA covariances; \textit{SS} means secondary structure predicted by the PETfold based on MSA; \textit{+} means a combination of features by a channel-wise concatenation.
        \item[c]  \textit{RNA-FM} means the RNA-FM embeddings. 
      \end{tablenotes}
\end{threeparttable}

\end{table}

Appendix Figure \ref{Fig.rnacontact}(a) shows the long-range Top-L precision distribution across all samples in TE80, comparing the ResNet32 with different input features. The y-axis of the plot represents RNA-FM embedding with transfer learning (\textit{RNA-FM(TL)}), and the x-axis represents the combination of MSA covariance and secondary structure predicted by the PETfold as input (\textit{Cov$+$SS}), which RNAcontact requires to generate from RNA MSA data. The RNA-FM with transfer learning matches or exceeds the MSA feature combination on \textbf{77.5\%} of instances of all RNA types in the form of most points over the diagonal. We also explore the relationship between the Top-L precision and the input RNA sequence length, as shown in Appendix Figure \ref{Fig.rnacontact}(b). The RNA-FM embedding with transfer learning outperforms the MSA features across all sequence lengths.

% Fig.\ref{Fig.rnacontact}(a) shows the long-range Top-L precision distribution across all samples in TE80, comparing the ResNet32 with different input features. The y-axis of the plot represents RNA-FM embedding (\textit{RNA-FM}), and the x-axis represents the combination of MSA covariance and secondary structure predicted by the PETfold as input (\textit{Cov$+$SS}), which RNAcontact requires to generate from RNA MSA data. The RNA-FM matches or exceeds the MSA feature combination on \textbf{61.3\%} of instances of all RNA types in the form of most points over the diagonal. We also explore the relationship between the Top-L precision and the input RNA sequence length, as shown in Fig.\ref{Fig.rnacontact}(b). The RNA-FM embedding outperforms the MSA features across all sequence lengths.

Appendix Figure \ref{Fig.rnacontact}(c) presents the predicted probability maps of two randomly selected examples in the TS0. With the standalone RNA-FM embedding (\textit{RNA-FM}, third column) as the input, the downstream model has already generated visualizations much closer to the ground truth (first column) than other features. Furthermore, we can perform far better than the other methods by applying above mentioned transfer learning (second column).

\begin{figure}[!htbp]
    \centering
    \includegraphics[width=0.74\textwidth]{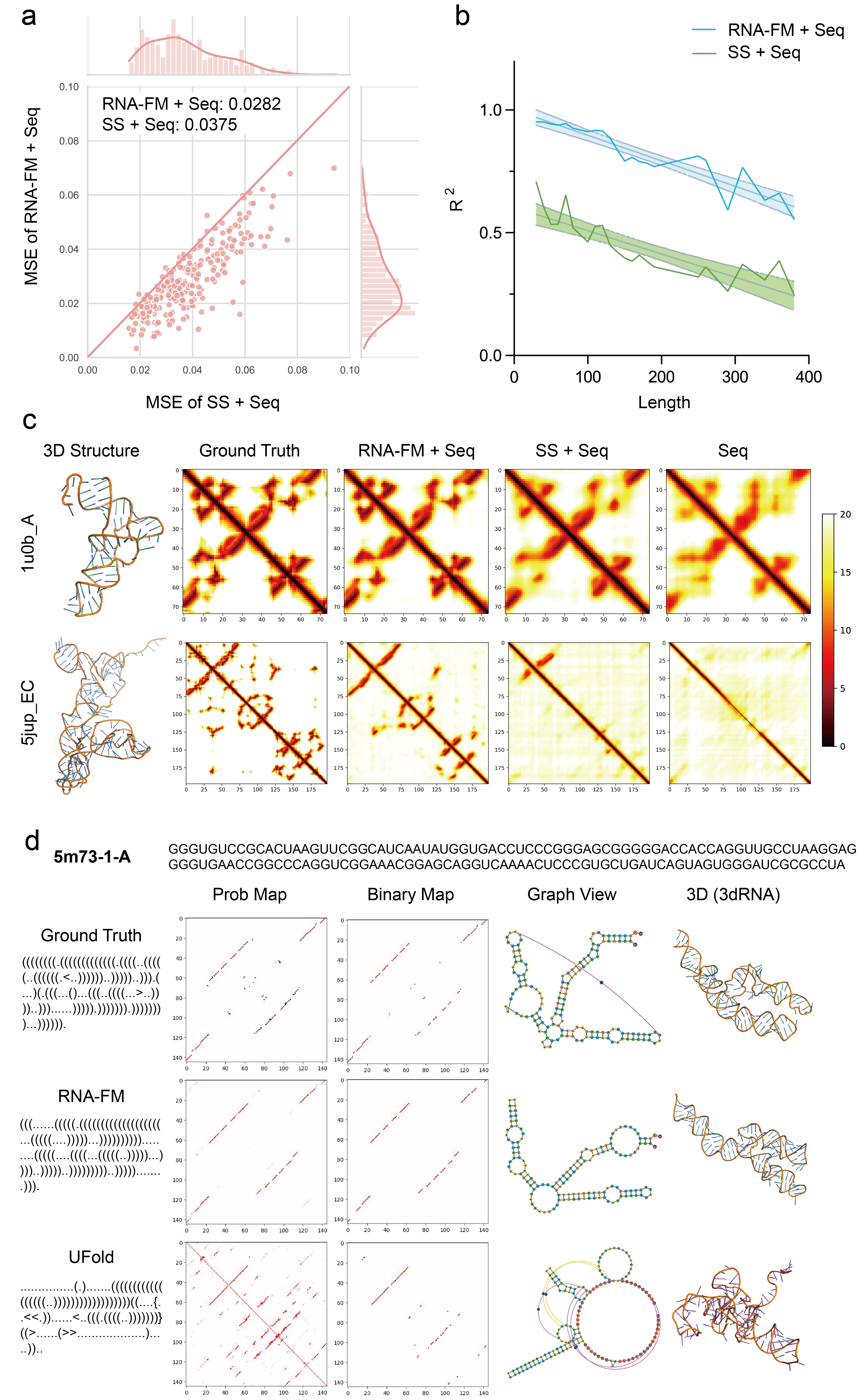}
    \caption{\textbf{RNA 3D distance prediction performance on the RNAcontact TE80 dataset and 3D reconstruction of RNA.} We use U-Net as the downstream base model with different input features. \textit{Seq}: sequence one-hot encoding; \textit{Cov} : MSA covariances; \textit{SS}: secondary structure from E2Efold; \textit{Emb}: embedding from RNA-FM; $+$: features combination via channel-wise concatenation. \textbf{a.} MSE Scatter plots with the performance of \textit{RNA-FM$+$Seq} as y-axis and \textit{SS$+$Seq} as x-axis. Each point represents an RNA structure. Almost all points are below the diagonal, which indicates that the RNA-FM embedding is superior over other features on nearly all instances. \textbf{b.} ${R^2}$ measurement as a function of RNA sequence lengths. \textit{RNA-FM$+$Seq} significantly outperforms that of \textit{SS$+$Seq} across all the lengths \textbf{c.} Probability maps of two randomly selected examples. Integrating RNA-FM embedding into input can significantly improve the model's performance. \textbf{d.} The probability maps and binary maps of an instance from PDB (5m73-1-A) are generated by different predictors. The graph views are obtained by jViz.Rna 4.0 \cite{shabash2017numerical}. Finally, the 3D structures are optimized and reconstructed by 3dRNA.}
    \label{fig:dist}
\end{figure}

%\subsubsection{RNA Distance Prediction}

\textbf{RNA distance map} defines the distance of arbitrary bases in the primary sequence. In the past few years, more and more complex protein structure prediction tasks have been thoroughly studied. For instance, trRosetta \cite{yang2020improved} can predict the distance between two amino acids and the orientation formed by their atom planes, and Alphafold \cite{alquraishi2019alphafold} can even directly predict the 3D structure of target proteins with high precision. However, 3D structure prediction in the RNA field is still an underdeveloped yet critical task. To ultimately approach this objective, we define a relatively new task for predicting this distance regression task, which can offer more information to downstream 3D folding methods than RNA secondary structure prediction \cite{singh2019rna} and 3D closeness prediction \cite{sun2021rna} mentioned above.

\begin{table}[!t]
\centering
\caption{\textbf{RNA 3D distance prediction performance on the RNAcontact TE80 dataset.} All the experiments are based on U-Net with different inputs. The model with standalone RNA-FM embedding can obtain a lower MSE than the model with sequence encoding and secondary structure information. When combined with sequence encoding, the RNA-FM embedding outperforms all the other feature combinations across different evaluation criteria. When combining sequence encoding, MSA covariances, and the RNA-FM embedding, we can reach the awe-inspiring performance of PMCC as high as 0.8313.}
\label{table:dist}
\begin{threeparttable}

\begin{tabular}{ccccc} 
\toprule
%Model &
Features & MSE   & R\textsuperscript{2} & PA(\%)   & PMCC    \\ 
\midrule
%\multirow{6}{*}{U-Net}  
Seq\tnote{a}             & 0.0615  & 0.3652  & 45.77  & 0.4024   \\
SS $+$ Seq        & 0.0387  & 0.6875  & 81.45  & 0.7826   \\
SS $+$ Cov $+$ Seq    & 0.0338  & 0.7821  & 85.44  & 0.8218   \\ 
\midrule  %\cmidrule(lr){1-5}
RNA-FM             & 0.0353  & 0.7542  & 84.62  & 0.8143   \\
RNA-FM $+$ Seq      & \underline{0.0322}  & \underline{0.7824}  & \underline{86.13}  & \underline{0.8261}   \\
RNA-FM $+$ Cov $+$ Seq   & \textbf{0.0319}  &
\textbf{0.7921} & \textbf{88.83}  & \textbf{0.8313}     \\
\bottomrule

\end{tabular}
\begin{tablenotes}
        \footnotesize
        \item[a]\textit{Seq} means sequences using one-hot encoding. \textit{Cov} means MSA covariances. \textit{SS} means secondary structures predicted by E2Efold. \textit{RNA-FM} means RNA-FM embedding. \textit{+} means a combination of features.
      \end{tablenotes}
\end{threeparttable}
\end{table}

%\textit{Evaluation} 
The dataset used for RNA distance prediction is the same as the benchmark used in the RNAcontact, as described in the previous paragraph. We generate distance maps for RNA sequences from their PDB files according to the minimal atomic distance of arbitrary bases. Then we limit the distance from 0 to 20 \r{A} and regard the value over 20 as 20. Finally, we use 20 to normalize the distance values and obtain a normalized distance map with elements falling into $[0,1]$.
Table \ref{table:dist} summarizes the distance prediction performance of the model with different inputs on TE80. The U-Net with RNA-FM embeddings as input significantly outperforms sequences across different evaluation criteria with almost a 39\% increase in $R^2$, a 39\% increase in Pixel Accuracy (PA), a 41\% increase in (PMCC), and 0.026 (42\%) decrease in MSE. Furthermore, when simply combining RNA-FM embeddings with sequences, we already obtained a model with a slightly better MSE over approaches that take advantage of all features, such as secondary structure (\textit{SS}) and MSA covariance (\textit{Cov}). It suggests that our generated RNA-FM embedding contains the most explicitly helpful information for this task. Notice that RNA-FM is a pure single-sequence method, eliminating the time-consuming MSA searching step. Moreover, when combining RNA-FM embeddings with sequences and MSA covariances, we obtain the best model with the lowest MSE at 0.0319. Evaluated $R^2$ also achieves the highest value of 43\% over the standalone sequence.
%1\% over the combination of other features (\textit{SS$+$Cov$+$Seq}.   %\yu{Highligh again that our method is a pure single-sequence method, which eliminates the time-consuming MSA searching step.}
%\paragraph{Downstream Module \& Training Scheme} 
%Considering the difficulty of this pairwise regression task, we build our downstream module upon classical U-Net architecture \cite{ronneberger2015u} with channels numbers of $[64,128,256,128,1026]$ for this task. We train the model with a batch size of $=1$, an optimizer of Adam, and a learning rate of $=0.001$. \cite{DBLP:journals/corr/KingmaB14}. We trained around $50$ epochs and selected the best validation results for each input feature. To evaluate the effectiveness of our embeddings, we try several input types, including sequences, RNA-FM embeddings, the secondary structures predicted by E2Efold, and their combinations. To match the dimension with other features, we also employ an MLP to map our embedding dimension from 640 to 128. Furthermore, the secondary structure features (SS) used in this section are obtained from E2Efold \cite{chen2020rna}.

% \begin{figure}[h!]
%     \centering
%     \includegraphics[width=1.0\textwidth]{figs/dist/dist.png}
%     \caption{Distance prediction examples.}
%     \label{fig:dist}
% \end{figure}
% \begin{figure}[h!]
%     \centering
%     \includegraphics[width=1.00\textwidth]{figs/dist/dist_len.png}
%     \caption{Distance prediction evaluation, MSE and R-squared.}
%     \label{fig:dist_metric}
% \end{figure}
%\paragraph{Results} 

Detailed analysis is conducted between two models with different input features. One is the combination of sequences and RNA-FM embeddings (\textit{RNA-FM$+$Seq}), and the other is the combination of sequences and the predicted secondary structures (\textit{SS$+$Seq}). Figure \ref{fig:dist}(a) shows the MSE of these two models across all instances in TE80, with \textit{RNA-FM$+$Seq} as the y-axis and \textit{SS$+$Seq} as the x-axis. The \textit{RNA-FM$+$Seq} is better than the \textit{SS$+$Seq} on \textbf{94.2\%} of instances in the form of most points below the diagonal. We also explore the relationship between the $R^2$ and the input RNA sequence length, as shown in Figure \ref{fig:dist}(b). When combined with sequence one-hot encoding, RNA-FM embedding outperforms the predicted secondary structure across all the RNA lengths.

As shown in Figure \ref{fig:dist}(c), we can see that our embedding feature enables the model to capture specific details of distance data while only providing sequential and secondary-structure data is not sufficient. The standalone sequential data does the worst in our experiment and only captures distance values on the diagonal.

\textbf{RNA 3D reconstruction} is the ultimate goal of RNA structure prediction. This section presents the results by combing RNA-FM with existing optimizing tools to obtain 3D approximates. Specifically, secondary structures are generated by the proposed RNA-FM and UFold. Then, their 3D structures can be optimized with RNA 3D modelling tools, including 3dRNA and FARFAR2 \cite{watkins2020farfar2}. We employ 3dRNA here to optimize the 3D structure upon its secondary structure.  
By applying our model to the PDB dataset \cite{singh2019rna}, we fine-tune the secondary structure predictor. The performance is shown in Supplementary Table 5. One of the examples in the PDB test set, namely TS1, is shown in Figure \ref{fig:dist}d. Notice that RNA-FM produces around $7.91$ RMSD, which is significantly better than the results produced by UFold ($25.70$). Interestingly, even the ground truth secondary structure produces a higher RMSD ($13.96$), suggesting that the error may actually come from the 3D structure modelling process. 
Moreover, we also apply it to the DCS-PK element in the 5' UTR flanking region of the Zika virus (ZIKV) \cite{li2018integrative}. The DCS-PK is a pseudoknot found in the coding region \cite{liu2013novel}, which helps enhance genome cyclization during replication. Due to the lack of ground truth for some 3D structures, RMSD for each prediction is unavailable, but RNA-FM produces more precise secondary structures than UFold for these targets, as shown in Appendix  Figure \ref{Fig.3dmodel-ren}.

To eliminate the deviation from the above 3D optimizing process, we also developed an end-to-end differentiable RNA 3D structures prediction model for evaluating the RNA-FM embedding by comparing them with the raw sequence inputs. 
As shown in Supplementary Table 6, on all the RNApuzzle structures, RNA-FM representations improve RNA 3D structure prediction greatly, with the average RMSD being around 4\AA{}.
It is consistent with all the above prediction tasks, suggesting that RNA-FM leads to more accurate RNA structure prediction.

\begin{figure}[htbp] %[!th]
\centering
\includegraphics[width=0.85\textwidth]{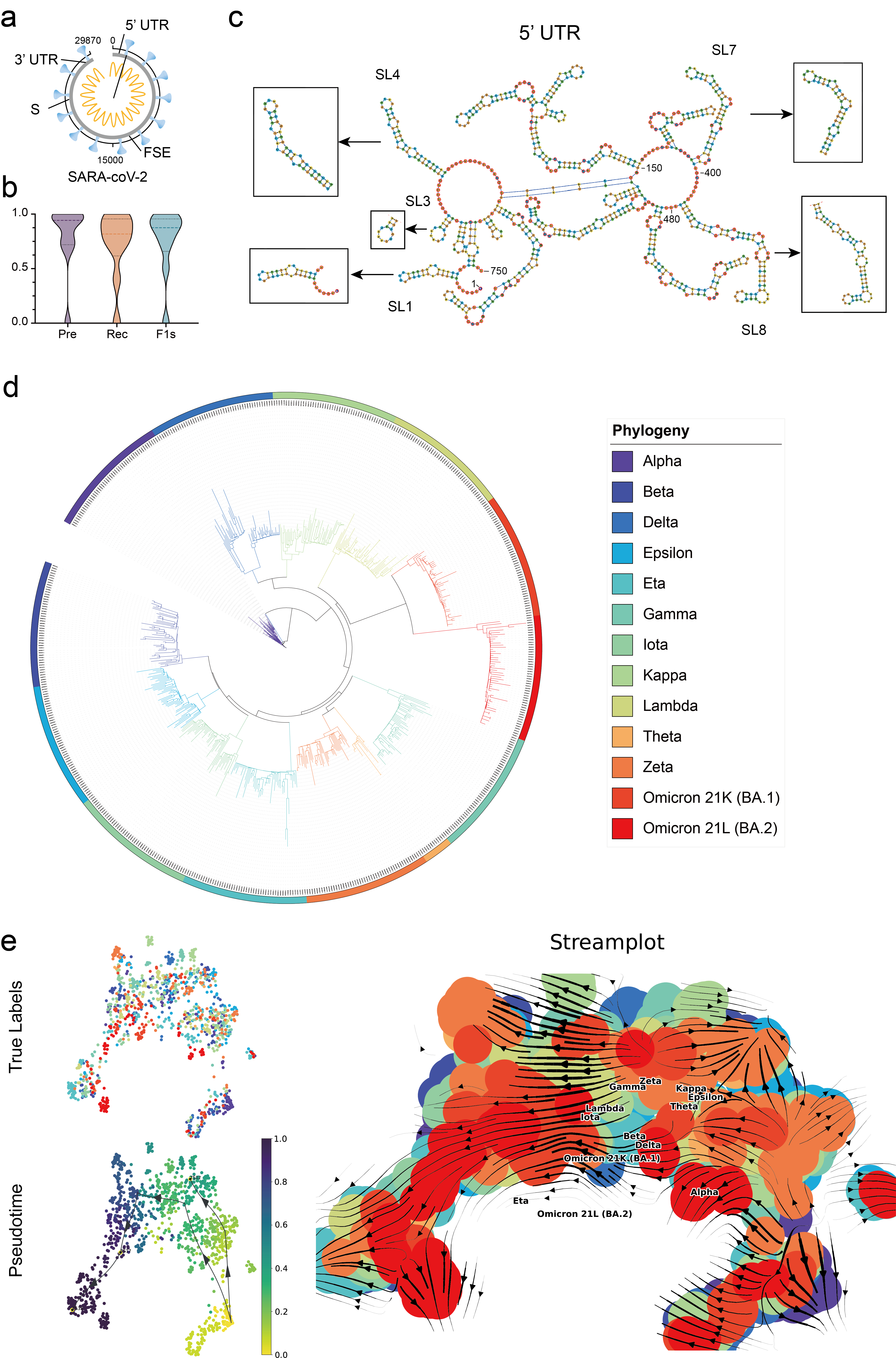} 
\caption{\textbf{RNA-FM predicts SARS-CoV-2 genome regulatory element secondary structures and the virus variant evolutionary trend.} \textbf{a.} Diagram of key regulatory segments sampled from the whole SARS-CoV-2 genome. \textbf{b.} Violin plots of the secondary structure prediction performance of segments mentioned above. RNA-FM can precisely predict the secondary structure of these segments. \textbf{c.} The visualization of RNA secondary structure predictions in 5'UTR. The prediction results are almost the same as the ground truth. \textbf{d.} The phylogenetic tree generated by FastME \cite{desper2002fast}. We treat it as the ground truth of the evolutionary trend of SARS-CoV-2, from Alpha types to Omicron variants. \textbf{e.} The trajectory inference of COVID-19 evolutionary trend with the RNA-FM genome-level embeddings. The results are highly consistent with the ground truth.}
\label{Fig.covid}
\end{figure}

%\siqi{there are only two targets, explain more details for both of them? RMSD info is not available in REN's lab target \jiayang{no ground truth} Ash: Jiayang, take a look}

\paragraph{RNA-FM facilities SARS-CoV-2 genome secondary structure and evolution study.}
COVID-19 has caused significant losses in properties and life in the past years, and detailed studies of the virus genome structure and its evolution are vital to prevent the next pandemic. We conduct such an investigation and apply RNA-FM to the whole genome of Severe acute respiratory syndrome coronavirus 2 (SARS-CoV-2), the causal pathogen of the epidemic \cite{wu2020new}.
Firstly, we utilize the well-trained RNA-FM to predict the secondary structures of the key regulatory segments of the SARS-CoV-2 reference genome (Refseq accession number NC\_045512.2). As shown in Figures \ref{Fig.covid}(a) and (b), we sample 3'UTR, 5'UTR, and other segments from the entire genome with the length of 29870 based on the work of Cao \textit{et al} \cite{cao2021architecture}. Our model precisely predicts the majority of them. The results indicate that our model can effectively perform RNA secondary structure prediction task on an independent test set, with desirable generalization property. The predictions of the fragments (in black boxes) in 5'UTR are visualized in Figure \ref{Fig.covid}(c), which are almost the same as the ground truth.
Secondly, we explore the evolution of different COVID variants by applying RNA-FM to the whole genomes. Although RNA-FM is not initially designed for whole genome modelling, we assume that aggregation of the RNA-FM embedding extracted from fragments of the whole genome can still characterize the genome, benefiting the study of the virus genome evolution. The calculation of genome-level RNA-FM embedding can be seen in the Method section. As shown in Figure \ref{Fig.covid}(e), the trajectory inference with the RNA-FM embedding of the virus genome is roughly in line with the phylogenetic tree generated by FastME \cite{desper2002fast}, which is considered the ground truth (Figure \ref{Fig.covid}(d)). The predictive evolution trend of COVID-19 begins with the Alpha type and ends up with the newest Omicron variant by April 2022, especially from Omicron 21K to Omicron 21L. 
Notice that RNA-FM is trained using merely the ncRNA sequences and unsupervised learning. We directly apply the trained model to the COVID-19 dataset without any fine-tuning or using any labelling information about the virus. It suggests that the regulatory elements of the virus genome could be vital for the virus variant evolution. Also, the RNA-FM framework can dig up core structure messages and evolutionary trend information of COVID-19 and its variants. Further development of the model has the potential to facilitate the research of COVID-19 and other pandemics.

%\paragraph{RNA Function Prediction }
%\paragraph{mRNA untranslated region's function prediction.}

%\begin{figure}[htbp]
%\centering
%\includegraphics[width=0.7\textwidth]{figs/UTR/UTRboxMAE.png} 
%\caption{examples} 
%\label{Fig.UTRboxMAE}
%\end{figure}

%\begin{figure}[htbp]
%\centering
%\includegraphics[width=0.7\textwidth]{figs/UTR/UTRboxMSE.png} 
%\caption{examples} 
%\label{Fig.UTRboxMSE}
%\end{figure}

%\subsection{RNA-protein Interaction Prediction}

% \yu{The logic of this application is not clear. We want to show that RNA-FM feature can achieve similar performnace as the in vivo RNA SS feature. Please point out this logic.}

\begin{table}[!t]
\centering
\caption{\textbf{ RBP-RNA binding prediction AUPRCs on the HeLa dataset with different RBPs.} CNN models with different input features are compared in this experiment. \textit{RNA-FM$+$Seq} and \textit{RealSS$+$Seq} outperform \textit{Seq} on all RBPs except for METTL14. Although the \textit{RealSS$+$Seq} achieves a higher mean AUPRCs than \textit{RNA-FM$+$Seq}, the latter surpasses the former on nearly half of RBPs, which demonstrates the effectiveness of the RNA-FM embedding.}
\label{RBP-tab}
\begin{threeparttable}

\begin{tabular}{cccc}
\toprule
RBPs          & Seq\tnote{a}         & RNA-FM $+$ Seq\tnote{b}   & RealSS $+$ Seq\tnote{c}   \\ 
\midrule
ELAVL1     & 0.946       & \textbf{0.950}          & 0.938       \\ 
UPF1       & 0.933       & 0.934         & \textbf{0.956}       \\
TIA1       & 0.917       & \textbf{0.929}         & 0.925       \\
METTL14    & \textbf{0.889}       & 0.866         & 0.876       \\
HNRNPC     & 0.874       & \textbf{0.883}         & 0.873       \\
CSTF2      & 0.860        & 0.867         & \textbf{0.870}        \\
TIAL1      & 0.853       & 0.852         & \textbf{0.869}       \\
U2AF65     & 0.852       & \textbf{0.899}         & 0.886       \\
eIF4AIII   & 0.818       & 0.814         & \textbf{0.829}       \\
WTAP       & 0.804       & 0.801         & \textbf{0.843}       \\
HNRNPU     & 0.798       & \textbf{0.829}         & 0.821       \\
PTBP1      & 0.793       & 0.799         & \textbf{0.840}        \\
PTBP1PTBP2 & 0.743       & 0.749         & \textbf{0.775}       \\
EIF4A3     & 0.738       & 0.751         & \textbf{0.758}       \\
U2AF2      & 0.712       & 0.705         & \textbf{0.764}       \\
METTL3     & 0.709       & \textbf{0.750}          & 0.704       \\
YTHDF2     & 0.610        & \textbf{0.638}         & 0.629       \\
\midrule
Mean    & 0.815 & \underline{0.824}   & \textbf{0.833} \\
\bottomrule
\end{tabular}
\begin{tablenotes}
        \footnotesize
        \item[a] \textit{Seq} means one-hot encoding of sequence (4 dims); \item[b] \textit{RNA-FM$+$Seq} combines the RNA-FM embedding (640 dims reduced to 1 dim) and the sequential one-hot encoding; \item[c] \textit{RealSS$+$Seq} means the combination of the structural score by the icSHAPE experiment with the sequence data.  %此处加入注释*信息
      \end{tablenotes}
\end{threeparttable}

\end{table}

% \begin{figure}[h!]
% \centering
% \subfigure[histogram]{
% \label{RBP1}
% \includegraphics[width=0.5\textwidth]{figs/PrismNet/Histogram.pdf}}
% \subfigure[violin plot]{
% \label{RBP2}
% \includegraphics[width=0.4\textwidth]{figs/PrismNet/violin.pdf}}
% \caption{Compared performance}
% \label{RBP}
% \end{figure}
\begin{figure}[!t]
    \centering
    \includegraphics[width=0.9\textwidth]{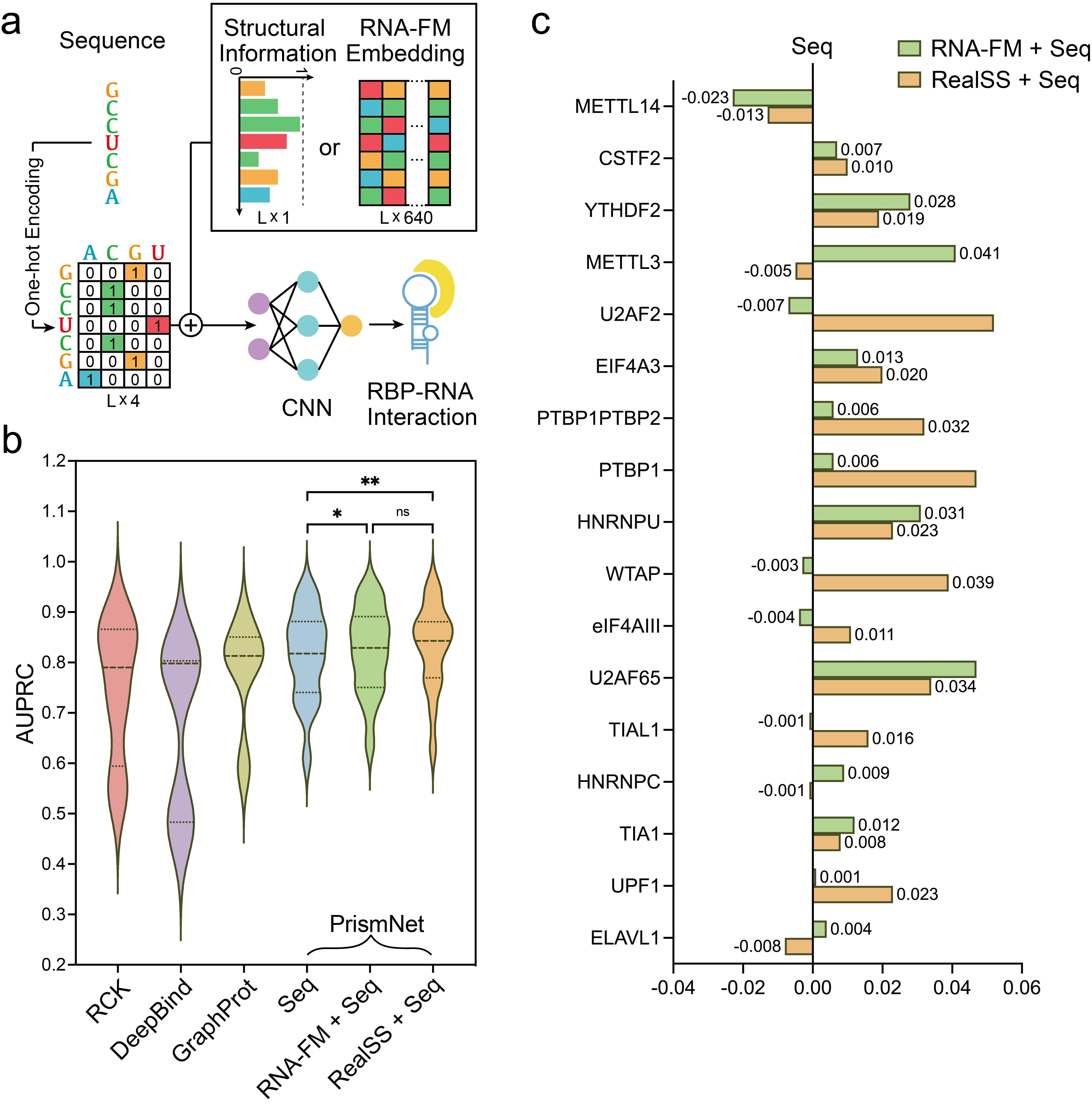}
    \caption{\textbf{RNA-FM embedding facilitates protein-RNA interaction prediction.} \textbf{a.} The deep learning framework for RBP-RNA interaction prediction in this experiment. The CNN model initially takes the combination of sequences and the structural scores offered by icSHAPE as input, while we substitute the experimental icSHAPE score with the RNA-FM embedding. \textbf{b.} Violin plots of AUPRCs of the model with different input features. Combining RNA-FM embedding with sequence encoding (\textit{RNA-FM$+$Seq}) can improve the median and the quartiles of AUPRCs over sequence standalone, approaching the performance generated by the experiment-measured structural information (\textit{RealSS$+$Seq}). \textbf{c.} Histogram plots of AUPRCs on different proteins with the \textit{Seq} as the baseline, corresponding to the vertical line across the origin point. The \textit{RNA-FM$+$Seq} outperforms the \textit{Seq} in most cases and sometimes achieves even better performance than the \textit{RealSS$+$Seq}, which shows the effectiveness of the RNA-FM features.}
    \label{fig:RBP}
\end{figure}

\paragraph{RNA-FM carries secondary structure information for RNA-protein interaction modelling.} Protein-RNA interactions are of vital importance in various cellular activities, including cell-signalling, post-transcriptional regulations, and protein synthesis \citep{wei2021protein}. We reproduce PrismNet \cite{sun2021predicting}, which includes \emph{in vivo} RNA secondary structure profiles for RNA-protein interaction prediction.  After that, we adopt RNA in the HeLa cell as the dataset for RNA binding protein prediction application and divide them into several sub-datasets according to different corresponding RBPs. The secondary structures of RNA are generated by icSHAPE (\emph{in vivo} click selective 2'-hydroxyl acylation and profiling experiment) \citep{spitale2015structural} in the HeLa cell environment. Then, we use our method to generate corresponding embeddings for all the sequences to replace the real secondary structures in PrismNet and make a comparison, evaluating the difference between the outputs from RNA-FM and the \emph{in vivo} secondary structure profiles.

For evaluation, we calculate AUPRCs on three streamlines (sequence only, sequence with real secondary structure, and sequence with RNA-FM) for comparison, as shown in Table \ref{RBP-tab} and Figure \ref{fig:RBP}(c). RNA-FM embeddings with sequences achieve the best performance on nearly half of the subsets. Their performance is even comparable to the real secondary structure with sequences, suggesting that embeddings from RNA-FM provide sufficient information as real secondary structures. Furthermore, taking advantage of RNA-FM embeddings helps the original model improve performance over models only with sequential information. Figure \ref{fig:RBP}(b) shows AUPRC violin plots for the three mentioned combinations as well as three other methods including RCK \citep{orenstein2016rck}, DeepBind \citep{alipanahi2015predicting}, GraphProt \citep{maticzka2014graphprot}. Our ``RNA-FM+Seq'' achieves close results compared to PrismNet with ``Real SS+Seq''. The RNA-protein interaction prediction results further illustrate that our embedding can learn sufficient information about secondary structures from raw sequences, which benefits the downstream functional prediction.

\begin{figure}[!t]
\centering
\includegraphics[width=0.9\textwidth]{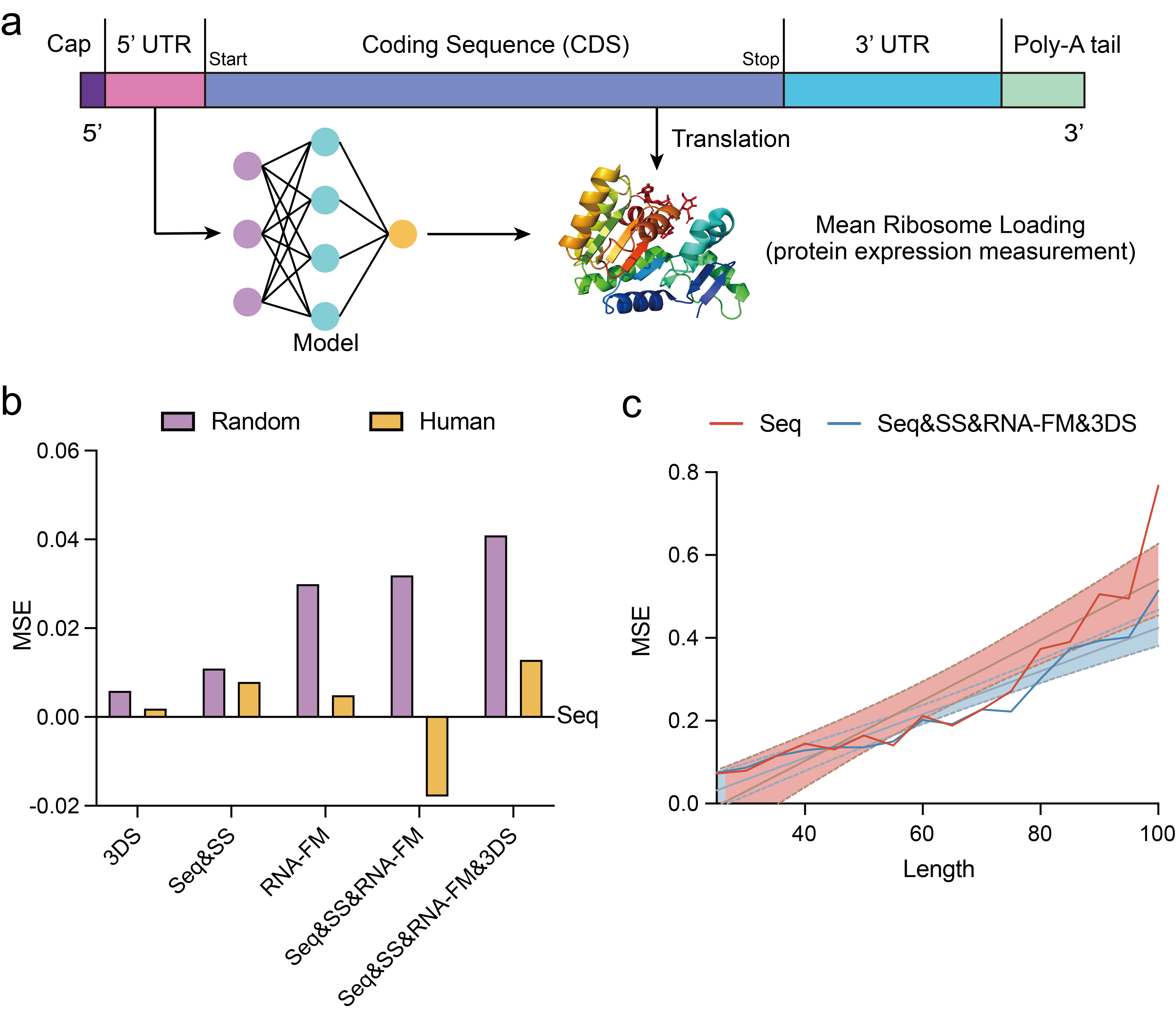}
\caption{\textbf{5’ UTR-based mean ribosome loading (MRL) prediction.} \textbf{a.} The deep learning framework for predicting MRL, which is a metric for evaluating the protein expression level regulated by UTR. \textbf{b.} Histogram plots of MSE on models of different input features with the one of \textit{Seq} as the baseline, corresponding to the horizontal line across the origin point. The \textit{RNA-FM} outperforms the \textit{Seq} significantly on the Random Set. When combined with the secondary structure (SS) and 3D structure (3DS) information, which is predicted based on RNA-FM, our method can further improve the performance on the Human Set. \textbf{c.} ${R^2}$ as a function of RNA sequence lengths. 
The model with information based on RNA-FM significantly outperforms \textit{Seq} model across all the lengths.
}
% \caption{\textbf{5’ UTR-based mean ribosome loading (MRL) prediction.} \textbf{a.} The deep learning framework for predicting MRL, which is a metric for evaluating the protein expression level regulated by UTR. 
% \textbf{b.} Violin plots of absolute error (the lower, the better) between the ground truth and predictions of the downstream model with different input features. \textit{Seq} means the one-hot encoding of the sequence; \textit{RNA-FM} means RNA-FM embedding; \textit{RNA-FM$+$Seq} means the channel-wise concatenation of these two features. \textit{RNA-FM} and \textit{RNA-FM$+$Seq} both obtain lower MSE and higher $R^2$ than \textit{Seq} on either the synthetic data (Random7600) or the real data (Human7600). \textbf{c.} ${R^2}$ as a function of RNA sequence lengths. \textit{RNA-FM} significantly outperforms \textit{Seq} across all the lengths. \textbf{d.} Prediction performance box plots grouped by different start codons context status. The \textit{RNA-FM$+$Seq} model performs slightly better than \textit{RNA-FM} in the context of none-uAUGs or OOF uURF regarding the absolute error, while \textit{RNA-FM} is better in the context of in-frame or out-frame uAUGs. \textbf{e.} Scatter plot of MRLs of all RNA samples with \textit{RNA-FM} predictions as the y-axis and the ground truth as the x-axis. The RNA sequences in the context of uAUG with lower MRLs than those without uAUG.
% }
\label{Fig.utr}
\end{figure}
\paragraph{RNA-FM generalizes to mRNA untranslated region’s function.} We further assume that RNA-FM could directly benefit the gene expression regulation modeling, which is one of the ultimate goals in the related studies, because function partially depends on structures. 
The 5’ untranslated region is the region of a messenger RNA (mRNA) located upstream of the initiation codon. This region is important for the translation regulation by different mechanisms in viruses, prokaryotes and eukaryotes. The sequence of 5’ UTR is a primary determinant of mRNA translation efficiency, especially the concomitant coding sequence (CDS), which is responsible for target protein expression. Although RNA-FM is trained with ncRNAs and the 5'UTR is a part of an mRNA (not belong to ncRNAs), we test the versatility of RNA-FM to generalize it on handling implicit non-coding sequences of an mRNA and aid with modelling the relationship between UTR and target protein expression.
With the assistance of massively parallel reporter assays and polysome profiling methods \cite{sample2019human}, which can measure the corresponding mean ribosome load (MRL) for each UTR, we can evaluate how a UTR regulates the target protein expression level of specific CDS by predicting the MRL of a UTR.
 
%  \jiayang{Although a 5'UTR is a part of a mRNA, which does not belong to ncRNAs, we attempt to test whether our pre-trained RNA-FM can handle these kinds of unseen non-coding sequences, and expect RNA-FM to finally aid with modeling the relationship between UTR and target protein expression.} %We adopt the pipeline and model as Paul et al. \cite{sample2019human} developed. The original inputs of their model are just the one-hot embeddings of sequences (4 dims). To see the effect of our embedding generated by RNA foundation model on this function-related task, we attempt to replace or add the sequences with our embeddings. Finally, we obtain three different types of inputs, including pure sequence (Seq) in the form of one-hot encoding (4 dims), pure RNA-FM embedding (RNA-FM) of 640 dims, and the combination of these two (RNA-FM + Seq). When the input includes embedding, we will apply a linear projection to reduce the embedding dimension from 640 to 4 for matching the one-hot embedding dimension.
%  \yu{Please incorporate Jiayang's point. The result is very impressive. Our method can generalize to non-nc regulatory RNA. It's very difficult. This point should be further highlighted.}

We utilize a large-scale synthetic Human 5’UTR library \cite{sample2019human} as the dataset for the UTR function prediction task. The dataset consists of 83,919 5’UTRs of 75 different lengths and their corresponding MRLs. 7600 sequences are sampled equally at each length as a validation set, while the remainder is adopted for training. An additional dataset consisting of 7600 real human 5’UTRs with the same length distribution provided by the library is used for validation to measure the generalization of models. %We adopt the pipeline and model as Paul et al. \cite{sample2019human} developed. The original inputs of their model are just the one-hot embeddings of sequences (4 dims). To see the effect of our embedding generated by RNA foundation model on this function-related task, we attempt to replace or add the sequences with our embeddings. Finally, we obtain three different types of inputs, including pure sequence (Seq) in the form of one-hot encoding (4 dims), pure RNA-FM embedding (RNA-FM) of 640 dims, and the combination of these two (RNA-FM + Seq). When the input includes embedding, we will apply a linear projection to reduce the embedding dimension from 640 to 4 for matching the one-hot embedding dimension.
We discover that model performances on the real human set are inferior to random set due to their data distribution difference, as illustrated in Table \ref{Tab.utr}. For both datasets, the model with RNA-FM embeddings is better than the model with pure sequences. On the synthetic set, model based on RNA-FM embedding can achieve $R^2$ = 0.875 and MSE = 0.247. On the human set, it  can achieve $R^2$ = 0.816 and MSE = 0.264. In addition, we assume that if we add more structure information to the model, we can further improve the modeling accuracy. So, we add the secondary structure information and 3D structure information, both predicted based on RNA-FM, into the model. As shown in Table \ref{Tab.utr} and \ref{Fig.utr}(b), the prediction accuracy is indeed further improved. 
% Embeddings generated by our model offer more explicit information valid for protein expression prediction than only pure sequences, which is shown in Figure \ref{Fig.utr}(b). 
Besides, performance gains are consistent across all the lengths and contexts, as shown in Figure \ref{Fig.utr}(c). This application further demonstrates the generalization of RNA-FM and its practical usage for real biological problems, even if the problem is not purely related to non-coding RNAs.

\begin{table}[!t]
\centering
\caption{\textbf{Mean ribosome loading (MRL) prediction performance on the Random7600 and Human7600 datasets.} Replacing the original input of MRL-CNN from sequence encoding (\textit{Seq}) to RNA-FM embedding (\textit{RNA-FM}) improves the performance.}
\label{Tab.utr}
\begin{threeparttable}

\begin{tabular}{ccccccc} 
\toprule
%Model  & 
\multirow{2}{*}{Features} &\multicolumn{3}{c}{Random7600} &\multicolumn{3}{c}{Human7600} \\
\cmidrule(r){2-4}\cmidrule(lr){5-7}
& R\textsuperscript{2} & MAE & MSE & R\textsuperscript{2} & MAE   & MSE \\ 
\midrule
%\multirow{2}{*}{MRL-CNN}   & 
Seq\tnote{a} &  0.860  & 0.371 & 0.277 &  0.814 & 0.375 & 0.269  \\ 
%\midrule
%\multirow{4}{*}{RNA-ESM} 
Seq $+$ SS\tnote{b} & 0.866 & 0.369 & 0.266 & 0.820& 0.370 & 0.261  \\ 
RNA-FM\tnote{c} & 0.876 & 0.360 & 0.247 & 0.816 & 0.377 & 0.264  \\ 
%\midrule
3DS\tnote{d} & 0.864 & 0.375 & 0.271 & 0.813 & 0.379 & 0.267 \\
Seq $+$ SS $+$ RNA-FM & 0.876 & 0.361 & 0.245 & 0.811 & 0.392 & 0.287  \\
Seq $+$ SS $+$ 3DS $+$ RNA-FM  & \textbf{0.882} & \textbf{0.353} & \textbf{0.236} & \textbf{0.824} & \textbf{0.368} & \textbf{0.256}  \\
\bottomrule
\end{tabular}

\begin{tablenotes}
        \footnotesize
        \item[a] \textit{Seq} means sequence encoding.
        \item[b] \textit{SS} means secondary structure, usually formatting an embedding (L$*$16) together with \textit{Seq}.
        \item[c] \textit{RNA-FM} means RNA-FM embedding.
        \item[d] \textit{3DS} means embedding extracted from 3D structure prediction framework.
      \end{tablenotes}
      
\end{threeparttable}

\end{table}
  
\section*{Discussion}
  % Future is bright~\cite{big}.
To take advantage of the abundant unannotated RNA data, we propose an RNA foundation model trained on 23 million RNA sequences via self-supervised learning, which can be employed in both structural and functional downstream applications. Detailed analysis shows that RNA-FM also encodes the evolutionary information implicitly, which can be used to derive the evolutionary trend of lncRNAs and SARS-CoV-2 variants. Several further experiments, from structure prediction to gene expression regulation modeling, are conducted, and the results strongly prove the effectiveness of our pre-trained model. Particularly in structural-related experiments, models which include our RNA-FM embeddings can significantly improve the performance among various tasks spreading from simple to complex. When dealing with a complex task with a relatively large-scale dataset, it is more likely to achieve admirable performance by fine-tuning our RNA-FM and downstream modules together. In the case of simple task with small-scale datasets, it is better to utilize transfer learning to avoid over-fitting. On all accounts, our RNA-FM indeed encodes the RNA structural patterns and can offer explicit information useful for RNA structure predictions.

However, the improvement brought by RNA-FM in the functional tasks seems more slight compared with the gain in the structural tasks. The underlying reason may be the sequence distribution differences between these function-related applications and our pre-training dataset. Besides, the relation between the RNA structure and its function is too complicated to represent directly. Even though without colossal performance improvement, our embedding can still be beneficial for these downstream tasks. We aim to provide more impressive results regarding these functional-related tasks in the future. %\yu{In fact, still impressive, although not that impressive than the structural part.}

\bibliographystyle{myrecomb}

\bibliography{mybib}

\begin{thebibliography}{10}
\expandafter\ifx\csname url\endcsname\relax
  \def\url#1{\texttt{#1}}\fi
\expandafter\ifx\csname urlprefix\endcsname\relax\def\urlprefix{URL }\fi
\providecommand{\bibinfo}[2]{#2}
\providecommand{\eprint}[2][]{\url{#2}}

\bibitem{miao2017rna}
\bibinfo{author}{Miao, Z.} \& \bibinfo{author}{Westhof, E.}
\newblock \bibinfo{title}{Rna structure: advances and assessment of 3d
  structure prediction}.
\newblock \emph{\bibinfo{journal}{Annual review of biophysics}}
  \textbf{\bibinfo{volume}{46}}, \bibinfo{pages}{483--503}
  (\bibinfo{year}{2017}).

\bibitem{caprara2000rna}
\bibinfo{author}{Caprara, M.~G.} \& \bibinfo{author}{Nilsen, T.~W.}
\newblock \bibinfo{title}{Rna: versatility in form and function}.
\newblock \emph{\bibinfo{journal}{Nature structural biology}}
  \textbf{\bibinfo{volume}{7}}, \bibinfo{pages}{831--833}
  (\bibinfo{year}{2000}).

\bibitem{atkins2011rna}
\bibinfo{author}{Atkins, J.~F.}, \bibinfo{author}{Gesteland, R.~F.} \&
  \bibinfo{author}{Cech, T.}
\newblock \bibinfo{title}{Rna worlds: from life's origins to diversity in gene
  regulation}  (\bibinfo{year}{2011}).

\bibitem{li2020strategies}
\bibinfo{author}{Li, B.}, \bibinfo{author}{Niu, Y.}, \bibinfo{author}{Ji, W.}
  \& \bibinfo{author}{Dong, Y.}
\newblock \bibinfo{title}{Strategies for the crispr-based therapeutics}.
\newblock \emph{\bibinfo{journal}{Trends in pharmacological sciences}}
  \textbf{\bibinfo{volume}{41}}, \bibinfo{pages}{55--65}
  (\bibinfo{year}{2020}).

\bibitem{bora2012rna}
\bibinfo{author}{Bora, R.~S.}, \bibinfo{author}{Gupta, D.},
  \bibinfo{author}{Mukkur, T. K.~S.} \& \bibinfo{author}{Saini, K.~S.}
\newblock \bibinfo{title}{Rna interference therapeutics for cancer: challenges
  and opportunities}.
\newblock \emph{\bibinfo{journal}{Molecular medicine reports}}
  \textbf{\bibinfo{volume}{6}}, \bibinfo{pages}{9--15} (\bibinfo{year}{2012}).

\bibitem{pardi2018mrna}
\bibinfo{author}{Pardi, N.}, \bibinfo{author}{Hogan, M.~J.},
  \bibinfo{author}{Porter, F.~W.} \& \bibinfo{author}{Weissman, D.}
\newblock \bibinfo{title}{mrna vaccines—a new era in vaccinology}.
\newblock \emph{\bibinfo{journal}{Nature reviews Drug discovery}}
  \textbf{\bibinfo{volume}{17}}, \bibinfo{pages}{261--279}
  (\bibinfo{year}{2018}).

\bibitem{chen2019computational}
\bibinfo{author}{Chen, X.} \emph{et~al.}
\newblock \bibinfo{title}{Computational models for lncrna function prediction
  and functional similarity calculation}.
\newblock \emph{\bibinfo{journal}{Briefings in functional genomicss}}
  \textbf{\bibinfo{volume}{18}}, \bibinfo{pages}{58--82}
  (\bibinfo{year}{2019}).

\bibitem{wang2011molecular}
\bibinfo{author}{Wang, K.~C.} \& \bibinfo{author}{Chang, H.~Y.}
\newblock \bibinfo{title}{Molecular mechanisms of long noncoding rnas}.
\newblock \emph{\bibinfo{journal}{Molecular cell}}
  \textbf{\bibinfo{volume}{43}}, \bibinfo{pages}{904--914}
  (\bibinfo{year}{2011}).

\bibitem{cech2014noncoding}
\bibinfo{author}{Cech, T.~R.} \& \bibinfo{author}{Steitz, J.~A.}
\newblock \bibinfo{title}{The noncoding rna revolution—trashing old rules to
  forge new ones}.
\newblock \emph{\bibinfo{journal}{Cell}} \textbf{\bibinfo{volume}{157}},
  \bibinfo{pages}{77--94} (\bibinfo{year}{2014}).

\bibitem{townshend2021geometric}
\bibinfo{author}{Townshend, R.~J.} \emph{et~al.}
\newblock \bibinfo{title}{Geometric deep learning of rna structure}.
\newblock \emph{\bibinfo{journal}{Science}} \textbf{\bibinfo{volume}{373}},
  \bibinfo{pages}{1047--1051} (\bibinfo{year}{2021}).

\bibitem{yao2019cellular}
\bibinfo{author}{Yao, R.-W.}, \bibinfo{author}{Wang, Y.} \&
  \bibinfo{author}{Chen, L.-L.}
\newblock \bibinfo{title}{Cellular functions of long noncoding rnas}.
\newblock \emph{\bibinfo{journal}{Nature cell biology}}
  \textbf{\bibinfo{volume}{21}}, \bibinfo{pages}{542--551}
  (\bibinfo{year}{2019}).

\bibitem{zuker1984rna}
\bibinfo{author}{Zuker, M.} \& \bibinfo{author}{Sankoff, D.}
\newblock \bibinfo{title}{Rna secondary structures and their prediction}.
\newblock \emph{\bibinfo{journal}{Bulletin of mathematical biology}}
  \textbf{\bibinfo{volume}{46}}, \bibinfo{pages}{591--621}
  (\bibinfo{year}{1984}).

\bibitem{waterman1978rna}
\bibinfo{author}{Waterman, M.~S.} \& \bibinfo{author}{Smith, T.~F.}
\newblock \bibinfo{title}{Rna secondary structure: A complete mathematical
  analysis}.
\newblock \emph{\bibinfo{journal}{Mathematical Biosciences}}
  \textbf{\bibinfo{volume}{42}}, \bibinfo{pages}{257--266}
  (\bibinfo{year}{1978}).

\bibitem{rivas2013four}
\bibinfo{author}{Rivas, E.}
\newblock \bibinfo{title}{The four ingredients of single-sequence rna secondary
  structure prediction. a unifying perspective}.
\newblock \emph{\bibinfo{journal}{RNA biology}} \textbf{\bibinfo{volume}{10}},
  \bibinfo{pages}{1185--1196} (\bibinfo{year}{2013}).

\bibitem{stadler2011viennarna}
\bibinfo{author}{Stadler, P.} \emph{et~al.}
\newblock \bibinfo{title}{Viennarna package 2.0}.
\newblock \emph{\bibinfo{journal}{Algorithms}}  (\bibinfo{year}{2011}).

\bibitem{hofacker1994fast}
\bibinfo{author}{Hofacker, I.~L.} \emph{et~al.}
\newblock \bibinfo{title}{Fast folding and comparison of rna secondary
  structures}.
\newblock \emph{\bibinfo{journal}{Monatshefte f{\"u}r Chemie/Chemical Monthly}}
  \textbf{\bibinfo{volume}{125}}, \bibinfo{pages}{167--188}
  (\bibinfo{year}{1994}).

\bibitem{markham2008unafold}
\bibinfo{author}{Markham, N.~R.} \& \bibinfo{author}{Zuker, M.}
\newblock \bibinfo{title}{Unafold}.
\newblock In \emph{\bibinfo{booktitle}{Bioinformatics}}, \bibinfo{pages}{3--31}
  (\bibinfo{publisher}{Springer}, \bibinfo{year}{2008}).

\bibitem{zuker2003mfold}
\bibinfo{author}{Zuker, M.}
\newblock \bibinfo{title}{Mfold web server for nucleic acid folding and
  hybridization prediction}.
\newblock \emph{\bibinfo{journal}{Nucleic acids research}}
  \textbf{\bibinfo{volume}{31}}, \bibinfo{pages}{3406--3415}
  (\bibinfo{year}{2003}).

\bibitem{huang2019linearfold}
\bibinfo{author}{Huang, L.} \emph{et~al.}
\newblock \bibinfo{title}{Linearfold: linear-time approximate rna folding by
  5'-to-3'dynamic programming and beam search}.
\newblock \emph{\bibinfo{journal}{Bioinformatics}}
  \textbf{\bibinfo{volume}{35}}, \bibinfo{pages}{i295--i304}
  (\bibinfo{year}{2019}).

\bibitem{mathews1998updated}
\bibinfo{author}{Mathews, D.~H.}, \bibinfo{author}{Andre, T.~C.},
  \bibinfo{author}{Kim, J.}, \bibinfo{author}{Turner, D.~H.} \&
  \bibinfo{author}{Zuker, M.}
\newblock \bibinfo{title}{An updated recursive algorithm for rna secondary
  structure prediction with improved thermodynamic parameters}
  (\bibinfo{year}{1998}).

\bibitem{reuter2010rnastructure}
\bibinfo{author}{Reuter, J.~S.} \& \bibinfo{author}{Mathews, D.~H.}
\newblock \bibinfo{title}{Rnastructure: software for rna secondary structure
  prediction and analysis}.
\newblock \emph{\bibinfo{journal}{BMC bioinformatics}}
  \textbf{\bibinfo{volume}{11}}, \bibinfo{pages}{1--9} (\bibinfo{year}{2010}).

\bibitem{lorenz2011viennarna}
\bibinfo{author}{Lorenz, R.} \emph{et~al.}
\newblock \bibinfo{title}{Viennarna package 2.0}.
\newblock \emph{\bibinfo{journal}{Algorithms for molecular biology}}
  \textbf{\bibinfo{volume}{6}}, \bibinfo{pages}{1--14} (\bibinfo{year}{2011}).

\bibitem{janssen2015rna}
\bibinfo{author}{Janssen, S.} \& \bibinfo{author}{Giegerich, R.}
\newblock \bibinfo{title}{The rna shapes studio}.
\newblock \emph{\bibinfo{journal}{Bioinformatics}}
  \textbf{\bibinfo{volume}{31}}, \bibinfo{pages}{423--425}
  (\bibinfo{year}{2015}).

\bibitem{zakov2011rich}
\bibinfo{author}{Zakov, S.}, \bibinfo{author}{Goldberg, Y.},
  \bibinfo{author}{Elhadad, M.} \& \bibinfo{author}{Ziv-Ukelson, M.}
\newblock \bibinfo{title}{Rich parameterization improves rna structure
  prediction}.
\newblock \emph{\bibinfo{journal}{Journal of Computational Biology}}
  \textbf{\bibinfo{volume}{18}}, \bibinfo{pages}{1525--1542}
  (\bibinfo{year}{2011}).

\bibitem{do2006contrafold}
\bibinfo{author}{Do, C.~B.}, \bibinfo{author}{Woods, D.~A.} \&
  \bibinfo{author}{Batzoglou, S.}
\newblock \bibinfo{title}{Contrafold: Rna secondary structure prediction
  without physics-based models}.
\newblock \emph{\bibinfo{journal}{Bioinformatics}}
  \textbf{\bibinfo{volume}{22}}, \bibinfo{pages}{e90--e98}
  (\bibinfo{year}{2006}).

\bibitem{sato2009centroidfold}
\bibinfo{author}{Sato, K.}, \bibinfo{author}{Hamada, M.},
  \bibinfo{author}{Asai, K.} \& \bibinfo{author}{Mituyama, T.}
\newblock \bibinfo{title}{Centroidfold: a web server for rna secondary
  structure prediction}.
\newblock \emph{\bibinfo{journal}{Nucleic acids research}}
  \textbf{\bibinfo{volume}{37}}, \bibinfo{pages}{W277--W280}
  (\bibinfo{year}{2009}).

\bibitem{nowakowski1997rna}
\bibinfo{author}{Nowakowski, J.} \& \bibinfo{author}{Tinoco~Jr, I.}
\newblock \bibinfo{title}{Rna structure and stability}.
\newblock In \emph{\bibinfo{booktitle}{Seminars in virology}},
  vol.~\bibinfo{volume}{8}, \bibinfo{pages}{153--165}
  (\bibinfo{organization}{Elsevier}, \bibinfo{year}{1997}).

\bibitem{singh2019rna}
\bibinfo{author}{Singh, J.}, \bibinfo{author}{Hanson, J.},
  \bibinfo{author}{Paliwal, K.} \& \bibinfo{author}{Zhou, Y.}
\newblock \bibinfo{title}{Rna secondary structure prediction using an ensemble
  of two-dimensional deep neural networks and transfer learning}.
\newblock \emph{\bibinfo{journal}{Nature communications}}
  \textbf{\bibinfo{volume}{10}}, \bibinfo{pages}{1--13} (\bibinfo{year}{2019}).

\bibitem{gutell2002accuracy}
\bibinfo{author}{Gutell, R.~R.}, \bibinfo{author}{Lee, J.~C.} \&
  \bibinfo{author}{Cannone, J.~J.}
\newblock \bibinfo{title}{The accuracy of ribosomal rna comparative structure
  models}.
\newblock \emph{\bibinfo{journal}{Current opinion in structural biology}}
  \textbf{\bibinfo{volume}{12}}, \bibinfo{pages}{301--310}
  (\bibinfo{year}{2002}).

\bibitem{chen2020rna}
\bibinfo{author}{Chen, X.}, \bibinfo{author}{Li, Y.}, \bibinfo{author}{Umarov,
  R.}, \bibinfo{author}{Gao, X.} \& \bibinfo{author}{Song, L.}
\newblock \bibinfo{title}{Rna secondary structure prediction by learning
  unrolled algorithms}.
\newblock \emph{\bibinfo{journal}{arXiv preprint arXiv:2002.05810}}
  (\bibinfo{year}{2020}).

\bibitem{sato2021rna}
\bibinfo{author}{Sato, K.}, \bibinfo{author}{Akiyama, M.} \&
  \bibinfo{author}{Sakakibara, Y.}
\newblock \bibinfo{title}{Rna secondary structure prediction using deep
  learning with thermodynamic integration}.
\newblock \emph{\bibinfo{journal}{Nature communications}}
  \textbf{\bibinfo{volume}{12}}, \bibinfo{pages}{1--9} (\bibinfo{year}{2021}).

\bibitem{fu2021ufold}
\bibinfo{author}{Fu, L.} \emph{et~al.}
\newblock \bibinfo{title}{Ufold: fast and accurate rna secondary structure
  prediction with deep learning}.
\newblock \emph{\bibinfo{journal}{bioRxiv}} \bibinfo{pages}{2020--08}
  (\bibinfo{year}{2021}).

\bibitem{xiong2021pairing}
\bibinfo{author}{Xiong, P.}, \bibinfo{author}{Wu, R.}, \bibinfo{author}{Zhan,
  J.} \& \bibinfo{author}{Zhou, Y.}
\newblock \bibinfo{title}{Pairing a high-resolution statistical potential with
  a nucleobase-centric sampling algorithm for improving rna model refinement}.
\newblock \emph{\bibinfo{journal}{Nature Communications}}
  \textbf{\bibinfo{volume}{12}}, \bibinfo{pages}{1--11} (\bibinfo{year}{2021}).

\bibitem{sun2021rna}
\bibinfo{author}{Sun, S.}, \bibinfo{author}{Wang, W.}, \bibinfo{author}{Peng,
  Z.} \& \bibinfo{author}{Yang, J.}
\newblock \bibinfo{title}{Rna inter-nucleotide 3d closeness prediction by deep
  residual neural networks}.
\newblock \emph{\bibinfo{journal}{Bioinformatics}}
  \textbf{\bibinfo{volume}{37}}, \bibinfo{pages}{1093--1098}
  (\bibinfo{year}{2021}).

\bibitem{lam2019deep}
\bibinfo{author}{Lam, J.~H.} \emph{et~al.}
\newblock \bibinfo{title}{A deep learning framework to predict binding
  preference of rna constituents on protein surface}.
\newblock \emph{\bibinfo{journal}{Nature communications}}
  \textbf{\bibinfo{volume}{10}}, \bibinfo{pages}{1--13} (\bibinfo{year}{2019}).

\bibitem{sun2021predicting}
\bibinfo{author}{Sun, L.} \emph{et~al.}
\newblock \bibinfo{title}{Predicting dynamic cellular protein--rna interactions
  by deep learning using in vivo rna structures}.
\newblock \emph{\bibinfo{journal}{Cell research}}
  \textbf{\bibinfo{volume}{31}}, \bibinfo{pages}{495--516}
  (\bibinfo{year}{2021}).

\bibitem{sample2019human}
\bibinfo{author}{Sample, P.~J.} \emph{et~al.}
\newblock \bibinfo{title}{Human 5' utr design and variant effect prediction
  from a massively parallel translation assay}.
\newblock \emph{\bibinfo{journal}{Nature biotechnology}}
  \textbf{\bibinfo{volume}{37}}, \bibinfo{pages}{803--809}
  (\bibinfo{year}{2019}).

\bibitem{devlin2018bert}
\bibinfo{author}{Devlin, J.}, \bibinfo{author}{Chang, M.-W.},
  \bibinfo{author}{Lee, K.} \& \bibinfo{author}{Toutanova, K.}
\newblock \bibinfo{title}{Bert: Pre-training of deep bidirectional transformers
  for language understanding}.
\newblock \emph{\bibinfo{journal}{arXiv preprint arXiv:1810.04805}}
  (\bibinfo{year}{2018}).

\bibitem{mcinnes2018umap}
\bibinfo{author}{McInnes, L.}, \bibinfo{author}{Healy, J.} \&
  \bibinfo{author}{Melville, J.}
\newblock \bibinfo{title}{Umap: Uniform manifold approximation and projection
  for dimension reduction}.
\newblock \emph{\bibinfo{journal}{arXiv preprint arXiv:1802.03426}}
  (\bibinfo{year}{2018}).

\bibitem{saelens2019comparison}
\bibinfo{author}{Saelens, W.}, \bibinfo{author}{Cannoodt, R.},
  \bibinfo{author}{Todorov, H.} \& \bibinfo{author}{Saeys, Y.}
\newblock \bibinfo{title}{A comparison of single-cell trajectory inference
  methods}.
\newblock \emph{\bibinfo{journal}{Nature biotechnology}}
  \textbf{\bibinfo{volume}{37}}, \bibinfo{pages}{547--554}
  (\bibinfo{year}{2019}).

\bibitem{necsulea2014evolution}
\bibinfo{author}{Necsulea, A.} \emph{et~al.}
\newblock \bibinfo{title}{The evolution of lncrna repertoires and expression
  patterns in tetrapods}.
\newblock \emph{\bibinfo{journal}{Nature}} \textbf{\bibinfo{volume}{505}},
  \bibinfo{pages}{635--640} (\bibinfo{year}{2014}).

\bibitem{stassen2021generalized}
\bibinfo{author}{Stassen, S.~V.}, \bibinfo{author}{Yip, G.~G.},
  \bibinfo{author}{Wong, K.~K.}, \bibinfo{author}{Ho, J.~W.} \&
  \bibinfo{author}{Tsia, K.~K.}
\newblock \bibinfo{title}{Generalized and scalable trajectory inference in
  single-cell omics data with via}.
\newblock \emph{\bibinfo{journal}{Nature communications}}
  \textbf{\bibinfo{volume}{12}}, \bibinfo{pages}{1--18} (\bibinfo{year}{2021}).

\bibitem{zhao2021review}
\bibinfo{author}{Zhao, Q.} \emph{et~al.}
\newblock \bibinfo{title}{Review of machine learning methods for rna secondary
  structure prediction}.
\newblock \emph{\bibinfo{journal}{PLoS computational biology}}
  \textbf{\bibinfo{volume}{17}}, \bibinfo{pages}{e1009291}
  (\bibinfo{year}{2021}).

\bibitem{tan2017turbofold}
\bibinfo{author}{Tan, Z.}, \bibinfo{author}{Fu, Y.}, \bibinfo{author}{Sharma,
  G.} \& \bibinfo{author}{Mathews, D.~H.}
\newblock \bibinfo{title}{Turbofold ii: Rna structural alignment and secondary
  structure prediction informed by multiple homologs}.
\newblock \emph{\bibinfo{journal}{Nucleic acids research}}
  \textbf{\bibinfo{volume}{45}}, \bibinfo{pages}{11570--11581}
  (\bibinfo{year}{2017}).

\bibitem{sloma2016exact}
\bibinfo{author}{Sloma, M.~F.} \& \bibinfo{author}{Mathews, D.~H.}
\newblock \bibinfo{title}{Exact calculation of loop formation probability
  identifies folding motifs in rna secondary structures}.
\newblock \emph{\bibinfo{journal}{RNA}} \textbf{\bibinfo{volume}{22}},
  \bibinfo{pages}{1808--1818} (\bibinfo{year}{2016}).

\bibitem{andronescu2003rnasoft}
\bibinfo{author}{Andronescu, M.}, \bibinfo{author}{Aguirre-Hernandez, R.},
  \bibinfo{author}{Condon, A.} \& \bibinfo{author}{Hoos, H.~H.}
\newblock \bibinfo{title}{Rnasoft: a suite of rna secondary structure
  prediction and design software tools}.
\newblock \emph{\bibinfo{journal}{Nucleic acids research}}
  \textbf{\bibinfo{volume}{31}}, \bibinfo{pages}{3416--3422}
  (\bibinfo{year}{2003}).

\bibitem{wayment2020rna}
\bibinfo{author}{Wayment-Steele, H.~K.}, \bibinfo{author}{Kladwang, W.},
  \bibinfo{author}{Participants, E.} \& \bibinfo{author}{Das, R.}
\newblock \bibinfo{title}{Rna secondary structure packages ranked and improved
  by high-throughput experiments}.
\newblock \emph{\bibinfo{journal}{BioRxiv}}  (\bibinfo{year}{2020}).

\bibitem{seemann2008unifying}
\bibinfo{author}{Seemann, S.~E.}, \bibinfo{author}{Gorodkin, J.} \&
  \bibinfo{author}{Backofen, R.}
\newblock \bibinfo{title}{Unifying evolutionary and thermodynamic information
  for rna folding of multiple alignments}.
\newblock \emph{\bibinfo{journal}{Nucleic Acids Research}}
  \textbf{\bibinfo{volume}{36}}, \bibinfo{pages}{6355--6362}
  (\bibinfo{year}{2008}).

\bibitem{wang2017accurate}
\bibinfo{author}{Wang, S.}, \bibinfo{author}{Sun, S.}, \bibinfo{author}{Li,
  Z.}, \bibinfo{author}{Zhang, R.} \& \bibinfo{author}{Xu, J.}
\newblock \bibinfo{title}{Accurate de novo prediction of protein contact map by
  ultra-deep learning model}.
\newblock \emph{\bibinfo{journal}{PLoS computational biology}}
  \textbf{\bibinfo{volume}{13}}, \bibinfo{pages}{e1005324}
  (\bibinfo{year}{2017}).

\bibitem{shabash2017numerical}
\bibinfo{author}{Shabash, B.} \& \bibinfo{author}{Wiese, K.~C.}
\newblock \bibinfo{title}{Numerical integration methods and layout improvements
  in the context of dynamic rna visualization}.
\newblock \emph{\bibinfo{journal}{BMC bioinformatics}}
  \textbf{\bibinfo{volume}{18}}, \bibinfo{pages}{1--18} (\bibinfo{year}{2017}).

\bibitem{yang2020improved}
\bibinfo{author}{Yang, J.} \emph{et~al.}
\newblock \bibinfo{title}{Improved protein structure prediction using predicted
  interresidue orientations}.
\newblock \emph{\bibinfo{journal}{Proceedings of the National Academy of
  Sciences}} \textbf{\bibinfo{volume}{117}}, \bibinfo{pages}{1496--1503}
  (\bibinfo{year}{2020}).

\bibitem{alquraishi2019alphafold}
\bibinfo{author}{AlQuraishi, M.}
\newblock \bibinfo{title}{Alphafold at casp13}.
\newblock \emph{\bibinfo{journal}{Bioinformatics}}
  \textbf{\bibinfo{volume}{35}}, \bibinfo{pages}{4862--4865}
  (\bibinfo{year}{2019}).

\bibitem{watkins2020farfar2}
\bibinfo{author}{Watkins, A.~M.}, \bibinfo{author}{Rangan, R.} \&
  \bibinfo{author}{Das, R.}
\newblock \bibinfo{title}{Farfar2: improved de novo rosetta prediction of
  complex global rna folds}.
\newblock \emph{\bibinfo{journal}{Structure}} \textbf{\bibinfo{volume}{28}},
  \bibinfo{pages}{963--976} (\bibinfo{year}{2020}).

\bibitem{li2018integrative}
\bibinfo{author}{Li, P.} \emph{et~al.}
\newblock \bibinfo{title}{Integrative analysis of zika virus genome rna
  structure reveals critical determinants of viral infectivity}.
\newblock \emph{\bibinfo{journal}{Cell host \& microbe}}
  \textbf{\bibinfo{volume}{24}}, \bibinfo{pages}{875--886}
  (\bibinfo{year}{2018}).

\bibitem{liu2013novel}
\bibinfo{author}{Liu, Z.-Y.} \emph{et~al.}
\newblock \bibinfo{title}{Novel cis-acting element within the capsid-coding
  region enhances flavivirus viral-rna replication by regulating genome
  cyclization}.
\newblock \emph{\bibinfo{journal}{Journal of virology}}
  \textbf{\bibinfo{volume}{87}}, \bibinfo{pages}{6804--6818}
  (\bibinfo{year}{2013}).

\bibitem{desper2002fast}
\bibinfo{author}{Desper, R.} \& \bibinfo{author}{Gascuel, O.}
\newblock \bibinfo{title}{Fast and accurate phylogeny reconstruction algorithms
  based on the minimum-evolution principle}.
\newblock In \emph{\bibinfo{booktitle}{International Workshop on Algorithms in
  Bioinformatics}}, \bibinfo{pages}{357--374}
  (\bibinfo{organization}{Springer}, \bibinfo{year}{2002}).

\bibitem{wu2020new}
\bibinfo{author}{Wu, F.} \emph{et~al.}
\newblock \bibinfo{title}{A new coronavirus associated with human respiratory
  disease in china}.
\newblock \emph{\bibinfo{journal}{Nature}} \textbf{\bibinfo{volume}{579}},
  \bibinfo{pages}{265--269} (\bibinfo{year}{2020}).

\bibitem{cao2021architecture}
\bibinfo{author}{Cao, C.} \emph{et~al.}
\newblock \bibinfo{title}{The architecture of the sars-cov-2 rna genome inside
  virion}.
\newblock \emph{\bibinfo{journal}{Nature communications}}
  \textbf{\bibinfo{volume}{12}}, \bibinfo{pages}{1--14} (\bibinfo{year}{2021}).

\bibitem{wei2021protein}
\bibinfo{author}{Wei, J.}, \bibinfo{author}{Chen, S.}, \bibinfo{author}{Zong,
  L.}, \bibinfo{author}{Gao, X.} \& \bibinfo{author}{Li, Y.}
\newblock \bibinfo{title}{Protein-rna interaction prediction with deep
  learning: structure matters.}
\newblock \emph{\bibinfo{journal}{Briefings in Bioinformatics}}
  (\bibinfo{year}{2021}).
\newblock \urlprefix\url{https://doi.org/10.1093/bib/bbab540}.

\bibitem{spitale2015structural}
\bibinfo{author}{Spitale, R.~C.} \emph{et~al.}
\newblock \bibinfo{title}{Structural imprints in vivo decode rna regulatory
  mechanisms}.
\newblock \emph{\bibinfo{journal}{Nature}} \textbf{\bibinfo{volume}{519}},
  \bibinfo{pages}{486--490} (\bibinfo{year}{2015}).

\bibitem{orenstein2016rck}
\bibinfo{author}{Orenstein, Y.}, \bibinfo{author}{Wang, Y.} \&
  \bibinfo{author}{Berger, B.}
\newblock \bibinfo{title}{Rck: accurate and efficient inference of sequence-and
  structure-based protein--rna binding models from rnacompete data}.
\newblock \emph{\bibinfo{journal}{Bioinformatics}}
  \textbf{\bibinfo{volume}{32}}, \bibinfo{pages}{i351--i359}
  (\bibinfo{year}{2016}).

\bibitem{alipanahi2015predicting}
\bibinfo{author}{Alipanahi, B.}, \bibinfo{author}{Delong, A.},
  \bibinfo{author}{Weirauch, M.~T.} \& \bibinfo{author}{Frey, B.~J.}
\newblock \bibinfo{title}{Predicting the sequence specificities of dna-and
  rna-binding proteins by deep learning}.
\newblock \emph{\bibinfo{journal}{Nature biotechnology}}
  \textbf{\bibinfo{volume}{33}}, \bibinfo{pages}{831--838}
  (\bibinfo{year}{2015}).

\bibitem{maticzka2014graphprot}
\bibinfo{author}{Maticzka, D.}, \bibinfo{author}{Lange, S.~J.},
  \bibinfo{author}{Costa, F.} \& \bibinfo{author}{Backofen, R.}
\newblock \bibinfo{title}{Graphprot: modeling binding preferences of
  rna-binding proteins}.
\newblock \emph{\bibinfo{journal}{Genome biology}}
  \textbf{\bibinfo{volume}{15}}, \bibinfo{pages}{1--18} (\bibinfo{year}{2014}).

\bibitem{vaswani2017attention}
\bibinfo{author}{Vaswani, A.} \emph{et~al.}
\newblock \bibinfo{title}{Attention is all you need}.
\newblock In \emph{\bibinfo{booktitle}{Advances in neural information
  processing systems}}, \bibinfo{pages}{5998--6008} (\bibinfo{year}{2017}).

\bibitem{dosovitskiy2020image}
\bibinfo{author}{Dosovitskiy, A.} \emph{et~al.}
\newblock \bibinfo{title}{An image is worth 16x16 words: Transformers for image
  recognition at scale}.
\newblock \emph{\bibinfo{journal}{arXiv preprint arXiv:2010.11929}}
  (\bibinfo{year}{2020}).

\bibitem{rives2021biological}
\bibinfo{author}{Rives, A.} \emph{et~al.}
\newblock \bibinfo{title}{Biological structure and function emerge from scaling
  unsupervised learning to 250 million protein sequences}.
\newblock \emph{\bibinfo{journal}{Proceedings of the National Academy of
  Sciences}} \textbf{\bibinfo{volume}{118}} (\bibinfo{year}{2021}).

\bibitem{rnacentral2021rnacentral}
\bibinfo{title}{Rnacentral 2021: secondary structure integration, improved
  sequence search and new member databases}.
\newblock \emph{\bibinfo{journal}{Nucleic acids research}}
  \textbf{\bibinfo{volume}{49}}, \bibinfo{pages}{D212--D220}
  (\bibinfo{year}{2021}).

\bibitem{fu2012cd}
\bibinfo{author}{Fu, L.}, \bibinfo{author}{Niu, B.}, \bibinfo{author}{Zhu, Z.},
  \bibinfo{author}{Wu, S.} \& \bibinfo{author}{Li, W.}
\newblock \bibinfo{title}{Cd-hit: accelerated for clustering the
  next-generation sequencing data}.
\newblock \emph{\bibinfo{journal}{Bioinformatics}}
  \textbf{\bibinfo{volume}{28}}, \bibinfo{pages}{3150--3152}
  (\bibinfo{year}{2012}).

\end{thebibliography}

\section*{Methods}
\paragraph{Overview of RNA-FM.}
In order to advance the development of RNA studies, we aim to build a unified foundation model in providing rich and meaningful representations inferred from standalone sequential information. We expect such representations to significantly boost various downstream tasks' performances when sufficient annotated data are unavailable. Inspired by recent great success in natural language processing, computer vision, and bioinformatics \cite{vaswani2017attention}\cite{dosovitskiy2020image}\cite{rives2021biological}, we explore the possibilities of a general transformer architecture in RNA-related studies. Thus, our framework is built upon the bidirectional transformer language model proposed in BERT \cite{devlin2018bert}, followed by the unsupervised training scheme. We named our framework RNA-FM, suggesting a foundational model for all the RNA-related studies.

This section first illustrates how we construct the large-scale ncRNA dataset, followed by model and training details. Next, we investigate a general strategy for several RNA function-related or structure-related downstream applications.
%Ash Jan-23

%Therefore, we use a Transformer \cite{vaswani2017attention, devlin2018bert, rives2021biological} model as backbone since it emerges as a powerful general model in multiple domains, such as natural language processing, computer vision and bioinformatics.\siqi{add some citation}. 

%In this section, we first present how we build the large-scale ncRNA dataset, then introduce the detailed model architecture and objection function. Finally, we propose a general framework for a set of RNA function-related or structure-related downstream tasks.

%\textcolor{blue}{%More importantly, considering that the obtaining of the annotations of RNA sequences is usually very costly in reality, we construct the RNA-FM method upon the unsupervised learning mechanism, which enables it to automatically learn the distribution and the sequential patterns and important structure information from the RNA sequences themselves without requiring any additional annotations.
%The proposed model can encode the distribution and the sequential pattern and functional information into its representations, and therefore contribute to a majority of structure-related and function-related tasks, which we named as RNA foundation model (RNA-FM).}\siqi{this part has been mentioned many times}

% 2.Dataset (Intro + Preprocessing)
\paragraph{Large-scale pre-training dataset.}

%\subsection{Large-scale Pre-training Dataset}
% \textcolor{blue}{Ash: Jan-24}
Our large-scale dataset used for pre-training phase is collected from RNAcentral \cite{rnacentral2021rnacentral}, the to-date largest dataset of the ncRNA. This dataset is indeed a comprehensive ncRNA sequences collection, representing all the ncRNA types from a broad range of organisms. It combines ncRNA sequences across 47 different databases, adding up to around 27 million RNA sequences in total.

We then pre-process all ncRNA sequences by replacing `T's with `U's since they are both complementary to adenine and similar in structure (`T's used for representing thymine in DNA while `U's stands for uracil in RNA.). This results in a dataset involving 4 main types of bases (16 counted types of combination in total, `A', `C', `G', `U', `R', `Y', `K', `M', `S', `W', `B', `D', `H', `V', `N', `-'). Moreover, to reduce redundancy without hurting the capacity of our dataset (reserve sequences as more as possible), we eliminate identical sequences by applying cd-hit-est \cite{fu2012cd} with a cut-off at 100\%.

After the above pre-processing steps, a final large-scale dataset consisting of 23.7 million ncRNA sequences is obtained. We named it RNAcentral100, which will be used to train our RNA foundation model in a self-supervised manner. Please refer to Supplementary Table.1-3 for the distribution of nucleotides and sequence lengths.
%Ash
%The pre-trained dataset of our foundation model is collected from RNAcentral, which is the to-date largest dataset of the non-coding RNA. The RNAcentral dataset is a comprehensive ncRNA sequence collection, representing all the ncRNA types from a broad range of organisms. This dataset combines ncRNA sequences from 47 different databases, including around 27 million RNA sequences in total.
%We pre-process all the sequences by replacing “T”s with “U”s\siqi{explain why or add a reference?}, which resulted a dataset with 20 types of amino acid in total. In addition, to preserve as many sequences as possible, we merely eliminate the identical sequences by applying cd-hit-est \cite{fu2012cd} with a cut-off at 100\%. After pre-processing, we obtain a large-scale dataset with ?\siqi{fill the number} million sequences, named RNAcentral100, which is used to train the proposed RNA foundation model. Please refer to table/figure \siqi{please refer table/figure number} for the distribution of 20 amino acids and sequence length.

% 3.RNA Foundation Model
\paragraph{RNA foundation model training details.}

% \textcolor{blue}{Ash: Jan-24}
Our RNA-FM's framework is a stack of $12$ transformer encoder blocks proposed in BERT \cite{devlin2018bert}\cite{vaswani2017attention}. Each encoder block \cite{devlin2018bert} consists of a $640$ hidden size feed-forward layer and a $20$ multi-heads self-attention layer. Layer normalization \cite{devlin2018bert} and residual connections are applied before and after every block, respectively.

For a RNA sequence with length $L$, RNA-FM takes raw sequential tokens as input, and an embedding layer maps each nucleotide token into a $640$-dimensional vector, thus resulting in an $L\times 640$ embedding matrix. The embedding matrix then proceeds through each encoder block, which includes multi-head self-attention modules and feed-forward layers. The output tensors from encoder blocks have the exact same size with the input, and a final Softmax layer is concatenated above to predict corresponding tokens including our selected 16 nucleotides and 4 specific functional identifiers. 

During the pre-training phase, we followed a self-supervised training manner in BERT \cite{devlin2018bert}. Around 15\% of nucleotide tokens are randomly replaced with a special mask token. ( If the $i$-th token is chosen, we replace the $i$-th token with (1) the [MASK] token 80\% of the time (2) a random token 10\% of the time (3) the unchanged $i$-th token 10\% of the time). We then train the model with masked language modeling (MLM) \cite{devlin2018bert} by predicting the original masked token with cross-entropy loss. Such a training strategy can be formulated as an objective function as follow: 

\begin{equation}
\mathcal{L}_{MLM}= \mathbb{E}_{x \sim \mathcal{X}}\mathbb{E}_{x_{\mathcal{M}}\sim x}\sum_{i \in \mathcal{M}} - \log p(x_i|x_{/\mathcal{M}}).  \label{EQ.1}
\end{equation}

Here in the above equation, a set of indices $\mathcal{M}$ are randomly sampled from each of the input sequences $x$ ($15\%$ among the whole sequence), and masked by replacing the true token at each index $i$ to some other mask tokens.
Next, for each masked token, when the masked sequence ($x_{/\mathcal{M}}$) is given as context, our adopted objective function will minimize the negative log-likelihood of the corresponding true nucleotide $x_i$.
Our adopted objective function can capture dependencies between masked proportion and the remaining parts of the input sequence, which enables accurate predictions for masked positions. Thus, RNA-FM trained via Eq.~(\ref{EQ.1}) drives the network to gain deep understanding and rich representation of each sequential token.

%To coordinate with the 26 million \siqi{the number in previous section is in question mark} sequences \siqi{coordinate means?}, we build the foundation model based on Transformer, the most prevalent architecture due to its high-capacity. The RNA-FM consists of 12 transformer layers with 640 embedding dimension and 20 attention heads. And it takes pure RNA sequences with nucleotide characters as input. More detailed, for a sequence $x$ with length $L$, we first tokenize each nucleotide from a character to an integer. Each token is then mapped into a 640-dimensional vector as the initial embedding. The embedding matrix then passes through the transformer layers, including a multi-head self-attention module and a feed-forward network (Please refer to appendix A for more details). The self-attention module explicitly incorporates the sequence embeddings from the previous layer using attention mechanism and eventually builds up a complex representation that models residue-residue interactions.

%In the pre-training stage, we apply the masked language modeling loss (MLM) \cite{devlin2018bert} to train the model. More specifically, 15\%  of the nucleotide are randomly replaced with a special mask token and RNA-FM aims to reconstruct these corrupted RNA sequences to the original ones. Therefore, RNA-FM is trained without using any additional annotated data. The objective function for MLM can be formulated as below:

%\siqi{the above equation seems wrong}

In summary, we trained RNA-FM on eight A100 GPUs of 80 GB memories for one month. We adopted an inverse square root learning rate schedule for the training of the neural network, with a 0.0001 base learning rate, a 0.01 weight decay, and 10000 warm-up steps. Similar to the previous studies, we also set the maximum length of the input sequences as 1024 to reduce the memory consumption and increase the batch size to speed up the training of the neural network.

\paragraph{RNA foundation model-generated embeddings analysis.}
After the training stage, we can get the embeddings for all the input RNA sequences generated by our RNA foundation model. In order to test if these embeddings have fully mined the structural and functional information belongs to RNA, we build RNA Atlas by subsample RNAcentral100 with a maximum of 10,000 samples per each RNA types. Each instance from different RNA families can be represented by a 640 dimensional vector by averaging across its RNA-FM embedding with shape of $L*640$ at each position in its sequence, where $L$ denotes the sequence length. Then, we apply UMAP \cite{mcinnes2018umap} to reduce these 640 dimensional vectors into 2 dimensional ones, and project them on a plane. Furthermore, we explore if the embeddings contain evolutionary information of RNA by applying trajectory inference, which is implemented by VIA \cite{stassen2021generalized}. We take embeddings of lncRNA as inputs, and get the stream-plot of RNA evolutionary trend.

% 4.Downtream Application 
\paragraph{Downstream training strategies.}
% \textcolor{blue}{Ash 13-Mar}
Once the pre-trained RNA-FM model encoded with RNA's informative structural and functional patterns is obtained, we can integrate it into various downstream applications via two schemes: feature-based training and fine-tuning. For the feature-based scheme, we freeze the parameters inside RNA-FM and feed them to downstream models by switching the output module. In contrast, the fine-tuning scheme will require training RNA-FM together with downstream modules rather than freezing parameters inside RNA-FM, which usually performs better than the feature-based training scheme.
% Regarding the downstream tasks fields that RNA-FM can handle, we think they mainly focus on structure-related tasks because the RNA primary structure has a very close relationship with its high-order structure. Because the structure also partially determines function, RNA-FM can deal with the function-related tasks to some extent.

Our proposed RNA-FM can effectively tackle both the RNA structure prediction and function modeling. We first investigate structure-related applications, including RNA secondary structure prediction, RNA 3D closeness prediction, and RNA distance prediction. 
% We expect to achieve SOTA on a broad field of evaluation metrics since RNA primary structure carries strong connections with its high-order structure. 
Additionally, we investigate the downstream function-related tasks, including COVID-19 virus modeling, RNA-protein interaction prediction, and gene expression regulation modeling. 
% Since RNA's primary structure can also partially determine its functions, RNA-FM is also expected to further boost its performance. 
We give more details about how to set up our methodology to perform in the major downstream applications.

\textbf{RNA second-order structure modeling.} 
% \textcolor{blue}{Ash} 
% \Ash{3-26}
RNA high-order structures can usually be represented by a 2D matrix. We would focus on three related studies to highlight the capability of RNA-FM: secondary structure, contact map, and distance map prediction. Secondary structure reflects its hydrogen bonds in primary sequence. While contact and distance maps concentrate more on pairwise tertiary inter-nucleotide interactions. Here, we adopt a simple 2D ResNet as a unified downstream prediction module for all the structure prediction tasks rather than creating an elaborately-designed framework for each sub-task. Similar to ESM-1b \cite{rives2021biological}, our deep residual network consists of 32 blocks, where each contains two convolution layers with a filter size of 64. The input of the ResNet32 is the outer concatenation of output embeddings obtained from query sequences. This module is utilized across all the RNA structure prediction tasks unless we specify (See Supplementary for more details of ResNet32). We show that RNA-FM can achieve significantly better performance than the state-of-the-art models with such simple downstream modules.

\textbf{End-to-end RNA 3D structure prediction.}
Similar to AlphaFold2 in protein research, we aim to first establish such an end-to-end differentiable model for RNA 3D structure prediction. The model takes advantage of 4 Evoformers as its backbone, and we stack an equivariant graph transformer (EGNN) on top as a 3D atom coordinate predictor. The data are collected via the up-to-date Protein Data Bank, including raw RNA sequences with corresponding 3D structures. 1036 sequences are used for training, and the left 100 are used for validation and testing purposes. We utilize distance RMSD as our objective for training, and we optimize the model for more than 10000 steps to get an initial version. To validate the effectiveness of our RNA-FM embeddings, we compare it with the pure RNA sequences as the inputs of the 3D framework.

\textbf{SARS-CoV-2 virus genome embedding extraction.} 
We aim to extract representations for severe acute respiratory syndrome coronavirus 2 (SARS-CoV-2) using RNA-FM. Considering the length of the whole genome (around 30K nucleotides) is far longer than the input limitation (1022 nucleotides) of RNA-FM, here, we apply a different feature extraction strategy. To be specific, we employ a fixed-length window of 1022nt on a non-overlap sub-section of the whole genome to extract the RNA-FM embeddings. Then, we aggregate embeddings by taking the average over them, and the final standard length vector is used as the RNA-FM embedding for the whole genome. We then apply trajectory inference with the genome-wise RNA-FM embeddings by VIA \cite{stassen2021generalized}. For each variant, we sample maximum 100 instances from the all sequences download from \href{https://www.ncbi.nlm.nih.gov/labs/virus/vssi/#/virus?SeqType_s=Nucleotide&VirusLineage_ss=Severe%20acute%20respiratory%20syndrome%20coronavirus%202,%20taxid:2697049&Completeness_s=complete}{SARS-CoV-2 Data Hub}.
We set $k$ in the $k$-nearest neighbors algorithm of VIA as 120 and the root as the Alpha variant.

\textbf{Protein-RNA interaction.} 
% \yu{We need three sentences to describe what is Protein-RNA interaction, why we want to study it, and what we want to predict. Also, what is the input and what is the output.} 
Protein-RNA interactions play important roles in a plenty of activities, such as cell-signaling, post-transcriptional regulations and protein synthesis \citep{wei2021protein}. Therefore, considering the importance of protein-RNA interactions in RNA function, we evaluate how well RNA-FM mine ncRNA function information by predicting RNA binding proteins corresponding to RNAs. In this application, we apply Hela cell as the dataset for the task and devide it into 17 different protein sub-sets. Within each one, we predict whether the input RNA can bind with the protein. Our downstream prediction module perform similarly as the PrismNet \cite{sun2021predicting} pipeline. Firstly, we reproduce PrismNet results, including two different types: One trained from raw sequences as baselines, and another one trained with \textit{in vivo} secondary structures. We then replace the actual secondary structures in PrismNet with our generated RNA-FM embeddings for all sequences while keeping the whole downstream PrismNet architecture unchanged. In this way, we can test if using RNA-FM embeddings can accurately predict protein-RNA interactions as the real RNA secondary structures.

\textbf{mRNA untranslated region's function.} 
% \textcolor{blue}{Ash} 
% \yu{We need three sentences to describe what is mRNA untranslated region, why we want to study it, and what we want to predict. Also, what is the input and what is the output.}
The 5' untranslated region is the region of a messenger RNA (mRNA). Although a 5'UTR is a part of a mRNA, which does not belong to ncRNAs, and RNA-FM is trained with ncRNAs, we attempt to test whether our pre-trained RNA-FM can handle these kinds of special non-coding sequences, and expect RNA-FM to finally aid with modeling the relationship between UTR and target protein expression. For the input UTR, we predict its corresponding mean ribosome load (MRL), which can reflect how a UTR regulates the target protein expression level. Specifically, we adopt the same pipeline and model as Paul et al. \cite{sample2019human} developed in order to testify the effectiveness of our RNA-FM embeddings. Their model is well-designed by performing a grid search, whose best framework is constructed as three 1D convolutional layers with 120 filters and a ReLU activation for each layer. The third convolution layer will output one channel and L length features, which will be fed into two fully-connected layers with one output node as the final prediction. The original inputs of the model are simply the one-hot representation of raw sequences (4 dims). Therefore, to stay as close to their original architecture as possible, we replace the inputs with our embeddings. Finally, we obtain three models with different inputs, including pure sequence (Seq) in the form of one-hot encoding (4 dims), pure RNA-FM embedding (RNA-FM) of 640 dims, and the combination of these two (RNA-FM + Seq). Further, we apply a linear projection to reduce the embedding dimension from 640 to 4 for matching the one-hot embedding dimension.

%\paragraph{RNA-protein Interaction Prediction} 
%Protein-RNA interactions play important roles in various cellular activities, such as posttranscriptional regulations and protein synthesis. Therefore, we also evaluate the performance of our designed network in the embeddings of a Protein-RNA interaction prediction pipeline. PrismNet is an effective tool to predict RBP-RNA binding sites based on the information from RNA sequence and secondary structures, which takes RNA sequences and in vivo secondary structures as inputs and can output a score for every binding site by evaluating every nucleotide position to determine whether it’s a binding site \citep{sun2021predicting}. Thus, we evaluate our method in PrismNet pipeline for the RNA-protein interaction prediction tasks.

\section*{Appendix}
    \renewcommand\figurename{Appendix Figure}

\begin{figure}[!t]
\centering
\includegraphics[width=0.95\textwidth]{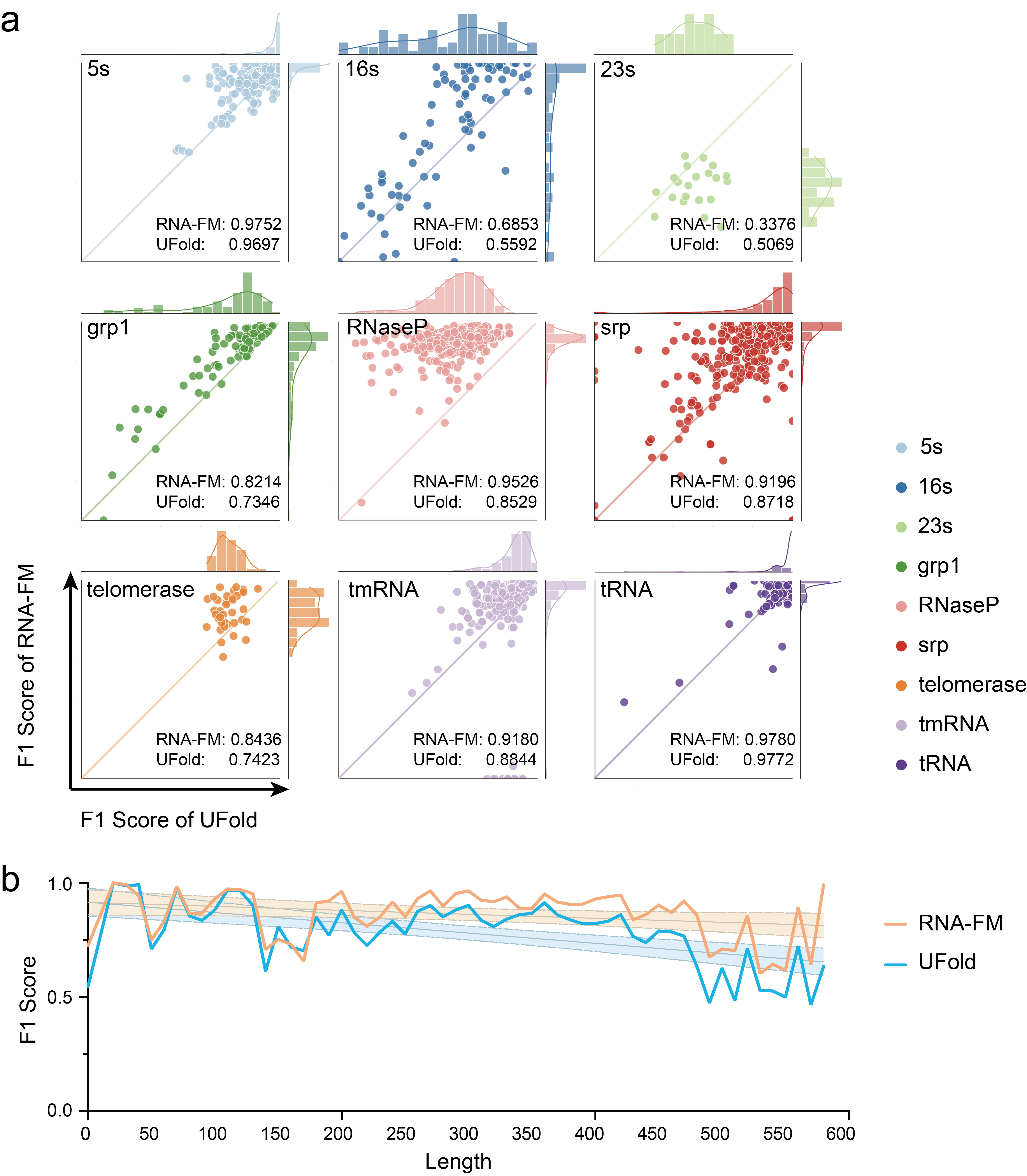} 
\caption{\textbf{Detailed performance comparison between RNA-FM and UFold on the ArchiveII dataset.} The experiment is cross-dataset validation of the trained model without re-training on ArchiveII. \textbf{a.} Scatter plots of F1 score comparison across 9 RNA types, with the performance of RNA-FM as the y-axis and that of UFold as the x-axis. Each point represents an RNA structure. Almost all the points are above the diagonal, which means RNA-FM beats UFold on nearly all the instances. \textbf{b.} F1 scores as a function of RNA sequence lengths. RNA-FM always outperforms UFold across all the lengths, especially when the length is over 150.} 
\label{Fig.UFold1}
\end{figure}

\begin{figure}[!th]
\centering
\includegraphics[width=0.95\textwidth]{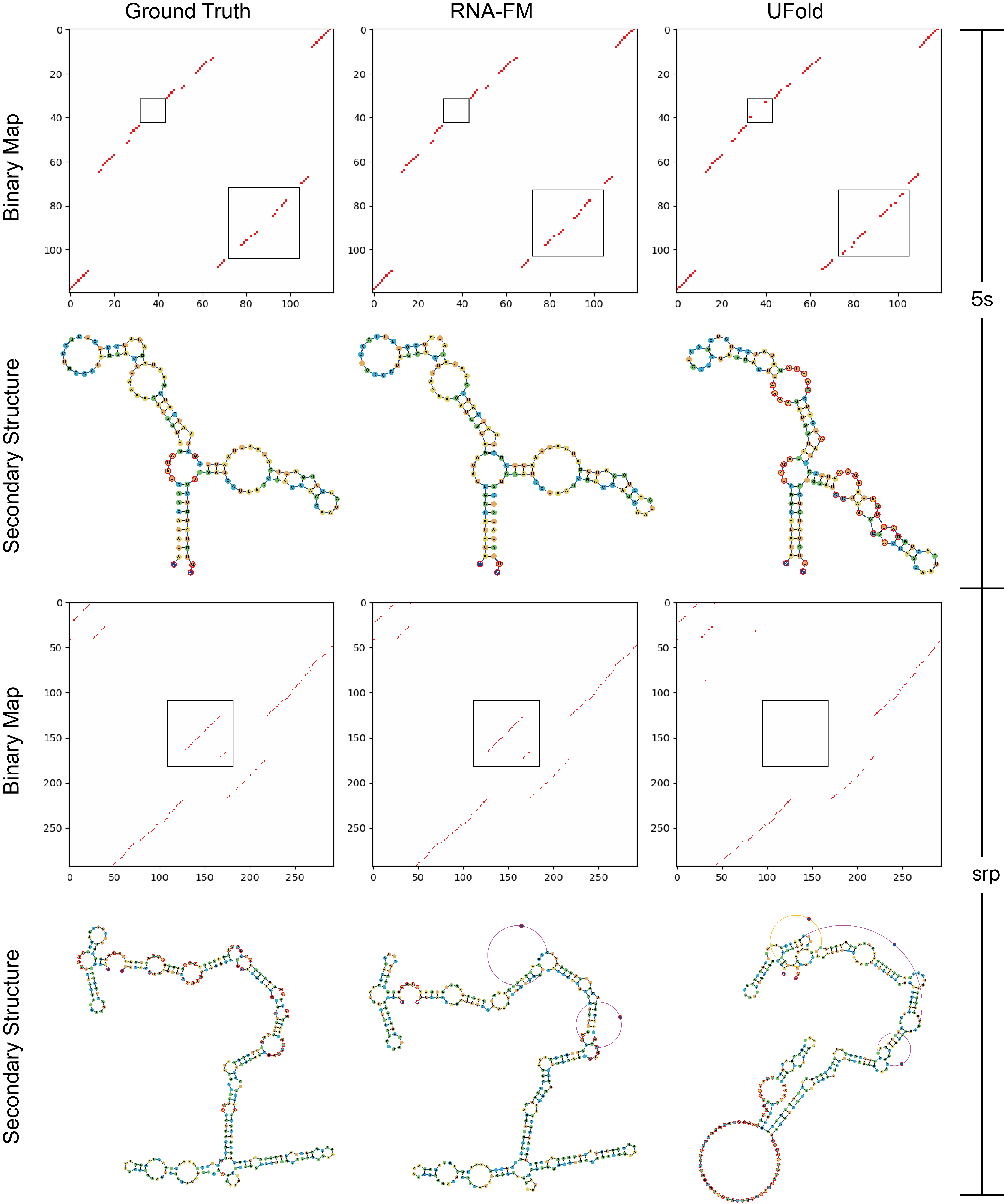} 
\caption{\textbf{Probability maps and graph view of secondary structure predictions of two randomly selected examples.} We compare the predictions from RNA-FM (second column) and UFold (third column) against the ground truth (first column). The probability maps from RNA-FM are more robust with less noise and closer to the ground truth compared to the ones from UFold. Regarding the visualization, RNA-FM also generates secondary structures more similar to the ground truth than UFold.}
\label{Fig.UFold2}
\end{figure}

\begin{figure}[!th]
\centering
\includegraphics[width=0.9\textwidth]{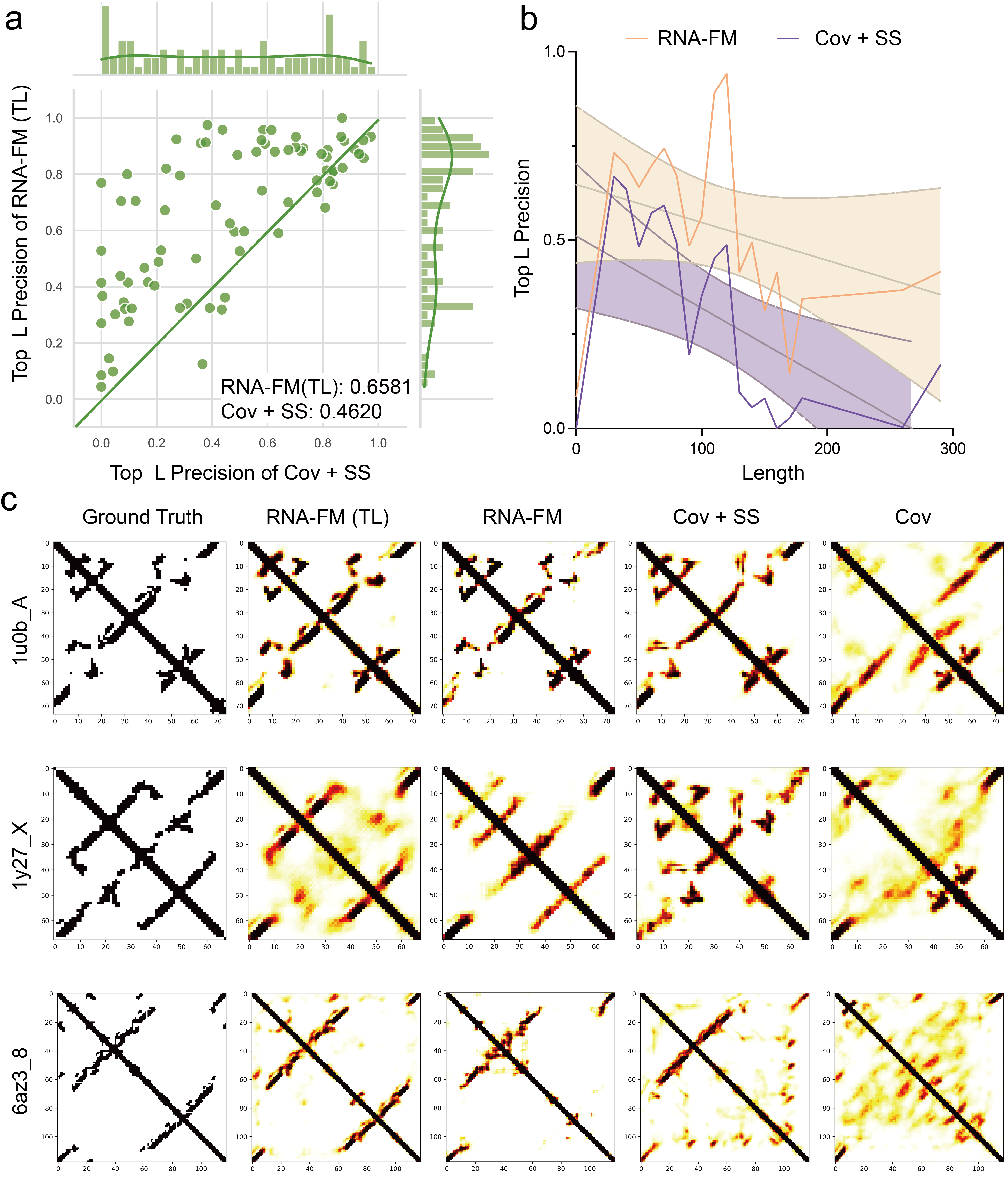}
\caption{\textbf{RNA 3D closeness prediction performance on RNAcontact TE80 dataset.} ResNet is used to reproduce RNAcontact results for an equal comparison of different features. \textit{Seq} means one-hot encoding of the sequence; \textit{Cov} means MSA covariances; \textit{SS} means secondary structure predicted by the PETfold based on MSA; \textit{RNA-FM} means the feature from RNA-FM;$+$ means a combination of features by a channel-wise concatenation. \textbf{a.} Scatter plots of MCCs, with the performance of \textit{RNA-FM} as the y-axis and the performance of \textit{Cov$+$SS} (used by RNAcontact) as the x-axis. Each point represents an RNA structure. Almost all the points are above the diagonal, which means RNA-FM embeddings are better than the other features in nearly all the instances. \textbf{b.} F1 scores as a function of RNA sequence lengths. \textit{RNA-FM} outperforms \textit{Cov$+$SS} all the time. \textbf{c.} Probability maps of three randomly selected examples. With the standalone RNA-FM embedding (\textit{RNA-FM}) as the input, the downstream model has already generated visualizations much closer to the ground truth than other features. Furthermore, by re-training the ResNet pre-trained on the other tasks using transfer learning, we can achieve the performance far better than the other methods.} 
\label{Fig.rnacontact}
\end{figure}

\begin{figure}[t] %[!th]
\centering
\includegraphics[width=1\textwidth]{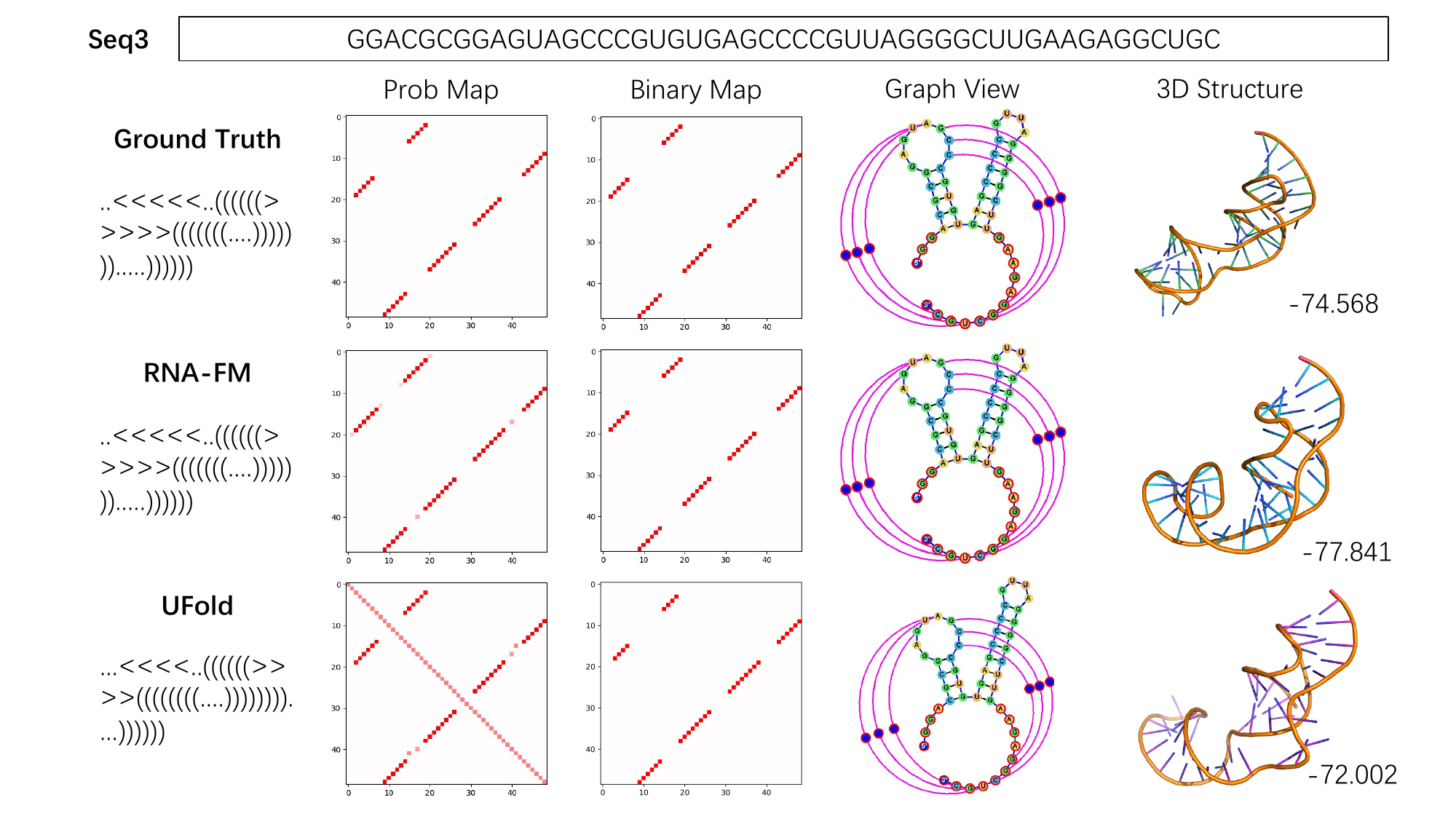} 
\caption{\textbf{3D reconstruction of RNA.} The probability maps and binary maps are generated by different predictors. The graph views are obtained by jViz.Rna 4.0 \cite{shabash2017numerical}. The 3D structures are modelled by 3dRNA. \textbf{a.} An instance from PDB (5m73-1-A). \textbf{b.} The DCS-PK of Zika Virus.}
\label{Fig.3dmodel-ren}
\end{figure}

\end{document}